\documentclass[arxiv]{ametsocV6.1}
\newcommand{\DR}{$\mathcal{D}_L(\mathbf{u})$}
\usepackage{array}
\usepackage{booktabs}
\usepackage{makecell}

\title{The representation of Convectively Coupled Equatorial Waves and upscale energy transfer in models with explicit and parametrized convection}

\authors{Elliot McKinnon-Gray,\aff{a}\correspondingauthor{Elliot McKinnon-Gray, e.b.mckinnon-gray@pgr.reading.ac.uk}
Daniel Shipley,\aff{a}
John Methven,\aff{a}
Thomas H. A. Frame,\aff{a}
Claudio Sanchez,\aff{b}
Anne McCabe,\aff{b}
and Nigel M. Roberts,\aff{c}}

\affiliation{\aff{a}{Department of Meteorology, University of Reading, Reading, UK}\\
\aff{b}{Met Office, Exeter, UK}\\
\aff{c}{European Centre for Medium-Range Weather Forecasts (ECMWF), Reading, UK}}

\abstract{Convectively Coupled Equatorial Waves (CCEWs) dominate atmospheric variability on timescales of 2–30 days in the Tropics, bringing episodes of widespread heavy precipitation. This study compares the representation of CCEWs and their connection to upscale energy transfer in two Met Office Unified Model simulations of the full tropical channel with identical km-scale resolution for the DYAMOND Summer period. The principal difference between the simulations is that one parametrizes convection (GAL9), while the other (RAL3) is convection permitting. The GAL9 convection scheme acts to remove vertical instability without explicitly representing the resolved-scale circulation associated with convective plumes. We present the first quantitative diagnosis of interscale energy transfer and its relation to CCEWs. This diagnosis is important because upscale energy transfer between convection and large-scale waves may influence accurate simulation of tropical weather systems. The average upper-tropospheric upscale transfer simulated by RAL3 is approximately 50\% higher than GAL9. CCEWs are more coherent in RAL3, with an average phase-speed variability 80\% higher than observations, compared with 166\% higher in GAL. RAL3 also simulates greater upscale energy transfer within waves than GAL9 with a stronger correlation between the interscale energy transfer and equatorial wave winds. Kelvin and Rossby waves are associated with upscale energy transfer from scales 4-8 times smaller than their wavelength, related to active deep convection within a particular sector of the wave phase. Our findings show that the explicit representation of convection has a significant impact on the simulation of upscale energy transfer, and is very likely to be a significant factor in the faithful simulation of convective coupling within. New text test hmm why not here
} 
\begin{document}
\maketitle

%
%
%
\statement
	 Weather forecasting in the tropics is beset by significant biases that are thought to be connected to a lack of scale coupling within current state-of-the-art numerical weather prediction models. This lack of scale coupling is in turn hypothesized to be due to an inability to explicitly represent the upscale transfer of energy from small to large scales. We hypothesize that the reason for this is due to the parametrization of convection in these models, and that kilometre-scale convection-permitting models will be able to simulate this upscale energy transfer and scale coupling, yielding improvements in simulation of the tropical atmosphere. We find that convection-permitting models simulate more upscale energy transfer and stronger scale coupling than conventional models with parametrized convection.
%


\section{Introduction}\label{s:Intro}

 The leading modes of variability on timescales from 2-30 days and spatial scales $\mathcal{O}(10^3)$ km in the tropical atmosphere are convectively coupled equatorial waves (CCEWs, \citealt{WK99,K09,Knippertz22}). CCEWs are equatorially-trapped propagating wave-like features that are coupled to propagating deep convection. There is a two-way interaction between the dynamical wave in terms of winds and geopotential height and deep convection that maintain the propagation of these systems. The behaviour of CCEWs is well known thanks to numerous observational studies \citep{WKW2000,SK03,Y03,y07p1}. Recent studies of CCEWs in global Numerical Weather Prediction (NWP) models with mass-flux parametrization schemes have found that current operational systems exhibit systematic flaws in the simulation of some CCEWs. For example, \citet{yang2021} found that Kelvin waves propagated too fast, and that there was a 25\% underestimate in Kelvin and westward mixed Rossby-gravity (WMRG) waves over the central Pacific. \citet{ferrett2023} found that skill in ensemble forecasts of Kelvin waves was substantially lower than for Rossby and WMRG waves over Southeast Asia. It has also been reported that CCEWs may be a very predictable component of the tropical atmosphere \citep{judt2020}. This would imply that better representation of CCEWs in models has significant potential to improve forecasting in the tropics.

Convection Permitting Models (CPMs) are one tool we can use that may yield improvements in the simulation of CCEWs. CPMs are NWP models that do not employ traditional convection parametrization schemes, instead opting for a finer grid spacing and allowing the model to explicitly simulate vertical convective motions and the exchanges of energy and momentum associated with moist convective instability. CPMs have provided benefits in many areas of tropical NWP, including improving the timing of the diurnal cycle of precipitation over the Maritime Continent \citep{howard2024}, improving short-range rainfall predictions over East Africa and Southeast Asia \citep{cafaro2021,ferrett2021}, improved propagation speed and amplitude of an active Madden-Julian Oscillation (MJO) event \citep{holloway2013}, improved vertical cloud structure of the West African Monsoon \citep{stein2015}, and more realistic propagation of African Easterly Waves (AEWs) \citep{birch2014}, to name some findings of the past decade. It is therefore reasonable to expect CPMs to improve the representation of CCEWs.

While operational NWP models are routinely run at convection permitting resolutions, this is done on limited area domains which do not wrap in longitude so limit the representation of zonally propagating CCEWs. In this study, we make use of the Met Office Unified Model (UM)'s Cyclic Tropical Channel (CTC) km-scale model. The CTC is a zonal global model with north and south boundaries at 26N and 40S to avoid the problem of grid-cell distortion approaching the poles. The main way we benefit from a CPM on this domain is that it makes the transform of the zonal coordinate to zonal wavenumber much easier as we are able to use Fourier transforms on this domain, which is not possible in a longitudinally bounded domain. The meridional extent is still limited, saving computing resources, but remains suitable for the purpose of studying equatorial waves.

A major difference between convection permitting and parametrized models is that CPMs represent near-gridscale circulations associated with convection, whereas these flows are absent in parametrized models. This means the upscale influence of convection on the larger scale motion is likely to be very different in the two models. We hypothesise that large-scale dynamical features $(L>10^3 \ \mathrm{km})$ receive energy from small scale processes (deep convection on km-scales, and aggregated convection on mesoscales $L\sim10^2 \ \mathrm{km}$) through a process known as upscale energy transfer. It is thought one of the key failings of NWP models regarding the simulation of large-scale dynamics has been an inability to properly simulate upscale energy transfer \citep{ShuttsPalmer2004,shutts2005kinetic}. We hypothesise that the primary reason for this is because the majority of current NWP models employ mass flux convection parametrization schemes, where vertical convective motions and the energy exchanges associated with them are parametrized at the grid scale by the model. Mass flux convection parametrization schemes are column-based and describe the effect of an ensemble of convective plumes on the large-scale environment which is assumed to be represented by the resolved fields in the model. Traditional convection parametrizations assume horizontal homogeneity, meaning that they cannot depend on horizontal gradients between grid cells. By design, the sub-grid components of ascent and descent within a column sum to zero at every level. Conversely, when convective systems are represented explicitly, there are horizontal motions coupling neighbouring cells linked with vertical motion, particularly in regions with large-scale vertical wind shear. This is seen, for example, in the structures identified with mesoscale convective systems \citep{dong2025MCS}. 

The most common methods of diagnosing energy transfer across scales \citep[e.g.][]{kolmogorov1941,NastromGage1985,Stephan2022} are spectral, and therefore inherently global in their formulation. One of the novelties of this study is that we will use a local formulation to diagnose energy transfer across scales, allowing us to analyse individual atmospheric processes associated with upscale energy transfer. In this study, we hypothesise that a CPM will capture the upscale transfer of energy from the scale of convective systems up to the scale of the waves in which they are embedded because of the explicit representation of vertical convective motions in the model. We also hypothesise that parametrized models will not be able to capture this transfer. It has also been noted \citep{selzcraig2015GRL,judt2018} that small-scale differences between simulations propagate upscale in CPMs, so a deeper understanding of upscale energy transfer is important for informing future development and use of CPMs. 

The outline of this paper is as follows. Section \ref{s:Theory} introduces the theory and formulation of the inter-scale energy transfer diagnostic. Readers who are not interested in the detailed fluid dynamics can proceed to the results comparing the simulations with and without convection parametrization (Section \ref{s:m&d}) and interscale energy transfer Section \ref{s:results&discuss}. Conclusions are outlined in section \ref{s:concs}.

\section{Local interscale energy transfers}\label{s:Theory}

Local interscale energy transfers describe the transfer of kinetic energy across a length scale $L$. For example, if $L=1000$ km, then the local interscale energy transfer determines whether there is a net transfer of energy from scales below 1000 km to larger than this or vice versa. By comparing the interscale energy transfer calculated for models that differ only in their treatment of convection we can isolate the upscale influence that this different treatment has on EWs. As a diagnostic of interscale energy transfer we use a method first developed for the atmosphere by \citet{Faranda2018}. The computation behind applying this method to high-resolution atmospheric data is described at the end of this section.

Much of the phenomenology used in this study is carried over from turbulence literature. In 3D homogeneous, isotropic turbulence, energy cascades downscale through interactions that are local in wavenumber and physical space. The power spectrum of such a ``forward" cascade is robustly predicted by the \citet{kolmogorov1941} spectrum. Turbulent flows that deviate from this behaviour are typically strongly anisotropic \citep{Augier2012}. They may be strongly stratified, dominated by layerwise horizontal motions, or the anisotropy might be associated with strong flow in one direction, such as a jet. To match this classical phenomenology of a forward cascade, an upscale transfer of energy is indicated by negative values, and a downscale transfer positive values (see figures). 

Diagnosing energy transfers across scales relies on the process of filtering (or coarse-graining) the velocity field to separate resolved motions from motions on scales smaller than the filter. The filtering approach was introduced by \citet{leonard1975}, and has been used widely in the turbulence and fluid mechanics communities \citep{germano1992,EyinkAluie2009,Green2020}. Recent examples of this technique applied to geophysical flows include \citet{Faranda2018,aluie2018JPO,storer2023,faranda2024}. In the traditional filtering approach used by \citet{aluie2018JPO} and \citet{EyinkAluie2009}, a filter is applied to the Navier-Stokes equations to highlight the contribution of the subfilter-scale stress tensor to the filtered (coarse-grained) velocities. From the coarse-grained equations, one can derive an energy balance for the kinetic energy of the filtered flow and study the contributions of the subfilter-scale energy transfers arising from nonlinear interactions \citep{meneveau1994}. 

An alternative formulation of the energy transfers in a fluid across a chosen scale $L$ was derived by \citet{DR2000}, and is the theoretical basis for the framework used to study local interscale energy transfers in the atmosphere by \citet{Faranda2018}. Using the weak formalism \citep{Leray1934} of the unforced incompressible Navier-Stokes equations, \citet{DR2000} showed that the kinetic energy obeys:
\begin{equation}
    \partial_t\left(\tfrac{1}{2}\mathbf{u}^2\right) 
  + \nabla \cdot 
    \left( \mathbf{u}\left(\tfrac{1}{2}\mathbf{u}^2 + p\right) \right) 
  - \nu \nabla^2 \tfrac{1}{2}\mathbf{u}^2 
  + \nu(\nabla \mathbf{u})^2 
  + \mathcal{D}(\mathbf{u}) 
    = 
    \mathbf{0},
    \label{eq:DR2000localenergy}
\end{equation}
where $\mathcal{D}(\mathbf{u})$ is an ``inertial dissipation'' that is nonzero only if the velocity field $\mathbf{u}$ is not smooth.

To derive this expression, Duchon and Robert introduce an infinitely differentiable, compactly-supported, non-negative function $\varphi$ normalised such that $\int \varphi = 1$ (called a ``test function'' in functional analysis and the theory of distributions). The test function $\varphi$ is used to smooth the weak solutions by integrating against them. In the filtering approach \citep{germano1992}, this represents the role of the filter, $G$. Filtering the incompressible Navier-Stokes equations, taking the inner product with the weak solution $\mathbf{u}$, adding the result to the inner product of the smooth solution with the un-smoothed Navier-Stokes equations, and simplifying, yields the following equation (equation~5 of \cite{Faranda2018}):
\begin{equation}
\begin{split}
    \underbrace{\partial_tE^L}_{\text{Energy tendency}}
    +
    \underbrace{\partial_j\Bigg(
        \underbrace{u_jE^L}_{\text{Advection}}
        + \frac{1}{2}\underbrace{(u_jp^L + u^L_jp)}_{\text{Pressure work}}
        + \frac{1}{4}\underbrace{\left( [u^2u_j]^L - [u^2]^L u_j \right)}_{\text{Subfilter-scale transport}}
        - \underbrace{\nu\partial_jE^L}_{\text{Viscous diffusion}}
    \Bigg)}_{\text{Divergence of fluxes (transport term)}}
    \\=
    -\underbrace{\nu\partial_j u_i\partial_j u_i^L}_{\text{Viscous dissipation}}
    - \underbrace{\mathcal{D}_L}_{\text{Local interscale energy transfer}} ,
\end{split}
    \label{eq:DRlocalenergybalance}
\end{equation}
where index notation (with the Einstein summation convention) is used, $u_i$ denotes the velocity field, $p$ is pressure divided by a constant reference density, $\nu$ is kinematic viscosity, superscript $L$ denotes coarse-grained quantities at scale $L$, and $E^L={{u_i^L}u_i}/{2}$, with the property that $\lim_{L\to0}E^L=u^2/2$. Although $u_i^Lu_i$ was interpreted by \cite{Faranda2018} as the kinetic energy per unit mass at scale $L$, it is best interpreted as a locally-averaged two-point velocity correlation function with characteristic length scale $L$; as such it can be both positive and negative.

The terms in (\ref{eq:DRlocalenergybalance}) are as follows. The first term on the LHS is the energy tendency. The second term on the LHS is the spatial transport term, which is comprised of a divergence of the spatial flux of energy $E^L$ (in brackets). The local interscale energy transfer term $\mathcal{D}_L$ is expressed in terms of velocity increments $\delta\mathbf{u}(\mathbf{r},\mathbf{x})=\mathbf{u}(\mathbf{x}+\mathbf{r})-\mathbf{u}(\mathbf{x})\equiv\delta\mathbf{u}(\mathbf{r})$ as
\begin{equation}
    \mathcal{D}_L(\mathbf{u}) = \frac{1}{4}\int \mathrm{d}\mathbf{r} (\nabla G_L)(\mathbf{r})\cdot\delta\mathbf{u}(\mathbf{r})|\delta\mathbf{u}(\mathbf{r})|^2,
    \label{eq:Dlu}
\end{equation}
where $\mathbf{x}$ is a position vector and $\mathbf{r}$ is a displacement vector.
Via an integration by parts this can be seen to be equal to a divergence \emph{in scale space}, making its interpretation as a transport in scale space unambiguous.
The same transfer term appears with opposite sign in the evolution equation for the quantity $\int \mathrm{d}\mathbf{r}\ G_L(\mathbf{r}) |\delta\mathbf{u}(\mathbf{r})|^2$, which is a characteristic kinetic energy of the small scales $< L$.\footnote{
    This evolution equation may be derived by integrating equation~(2.13) of \cite{Hill2002} against the kernel $G_L$.
}

Fundamental turbulence research typically considers unstratified 3D flows in Euclidean space far away from boundaries; hence spherically-symmetric filter kernels are usually employed. However, in the geosciences, the importance of gravity singles out the vertical direction as special. Therefore we consider only 2D filter kernels acting on surfaces perpendicular to gravity. We assume that these surfaces are concentric to the Earth's surface, which will itself be assumed to be spherical. Here our approach differs from \cite{Faranda2018}, who use a 3D kernel; Appendix~B provides further details of the differences between our approach and theirs.

On a curved surface $S$, the local interscale energy transfer term is written:
\begin{equation}
    \mathcal{D}_L(\mathbf{u}) = \frac{1}{4}\int_S \mathrm{d}S\ \left[\nabla G_L (r,\chi) \right] \cdot \delta\mathbf{u}(r,\chi)|\delta\mathbf{u}(r,\chi)|^2,
    \label{eq:Dlu_manifold}
\end{equation}
where $\delta\mathbf{u}(r,\chi) := \mathbf{u}(x_2) - \mathbf{u}(x_1)$, with $x_1$, $x_2$ two points on the surface $S$ separated by a geodesic distance $r$ at an angle $\chi \in [0, 2\pi]$ from an arbitrary reference direction.
For a sphere of radius $R$ (denoted $S^2(R)$), the surface element $\mathrm{d}S$ may be written:
\begin{equation}
    \mathrm{d}S =
    R \sin \alpha \mathrm{d}\chi R \mathrm{d}\alpha
    = R \sin\left(\frac{r}{R}\right) \mathrm{d}\chi \mathrm{d}r,
    \label{eq:dS}
\end{equation}
where $r$ is the great circle distance between $x_1$ and $x_2$, $\alpha \in (0,\pi]$ is the angle subtended by the surface of the sphere between $x_1$ and $x_2$, and $\chi$ the angle between the great circle arc connecting $x_1$ and $x_2$ and the line of constant latitude passing through $x_1$. For $\frac{r}{R} \ll 1$ this reduces to $\mathrm{d}S \approx r \mathrm{d}\chi \mathrm{d}r$, matching the surface element of Euclidean space expressed in polar coordinates. 

\begin{figure}[t]
 \noindent\includegraphics[width=39pc]{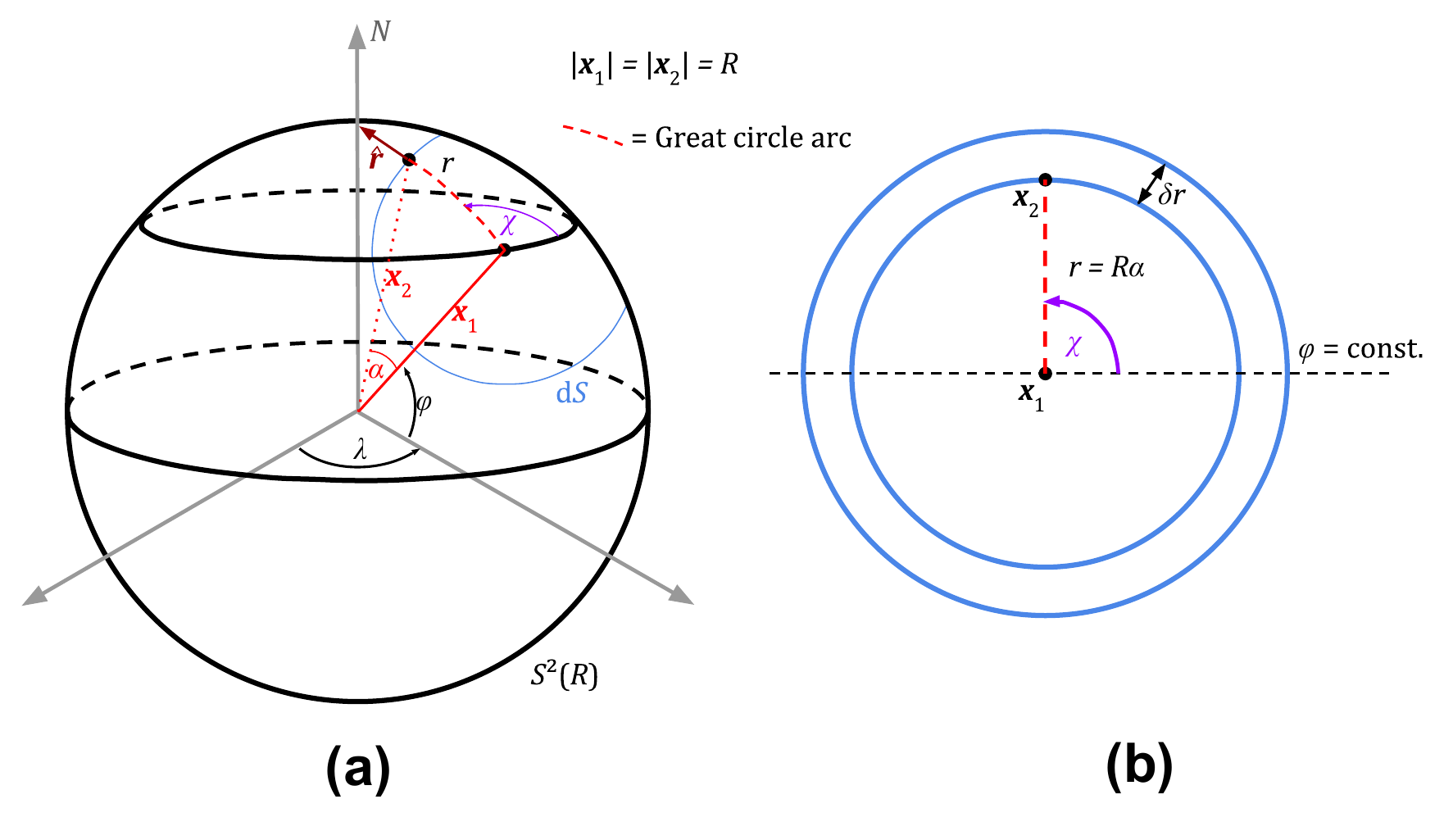}\\
 \caption{Schematic of the spherical coordinate system and elements described in equations (\ref{eq:dS}) and (\ref{eq:int_piphi}).}\label{fig:dS}
\end{figure}

For any 2D kernel that depends only on the great circle distance $r$, and not on the angle $\chi$, integrating over a spherical surface, (\ref{eq:Dlu}) becomes:
\begin{align}
    \int_{S^2(R)} \nabla G_L(r) \cdot \delta \mathbf{u} |\delta \mathbf{u}|^2 \mathrm{d}S
    &=
    \int_0^{\pi R} \int_0^{2 \pi}
        \frac{\partial G_L(r)}{\partial r} \hat{\mathbf{r}} 
      \cdot \delta \mathbf{u} |\delta \mathbf{u}|^2 R \sin\left(\frac{r}{R}\right) 
    \mathrm{d}\chi\ \mathrm{d}r
    \nonumber
    \\
    &=
    R 
    \int_0^{\pi R} 
    \frac{\partial G_L(r)}{\partial r} \sin\left(\frac{r}{R}\right) 
    \left[
        \int_0^{2 \pi}
            \hat{\mathbf{r}} 
          \cdot \delta \mathbf{u} |\delta \mathbf{u}|^2
        \mathrm{d}\chi 
    \right]
    \mathrm{d}r.
    \label{eq:int_piphi}
\end{align}
Here $\hat{\mathbf{r}}$ is the tangent vector at $x_2$ to the great circle arc passing through $x_1,x_2$. A schematic of the quantities in equations (\ref{eq:dS}) and (\ref{eq:int_piphi}) can be found in Fig. \ref{fig:dS}.

Since the sphere is compact, it only makes sense to consider compactly-supported filter kernels. One such compactly supported filter kernel is the ``standard mollifier" used by \citet{kuzzay2015,kuzzay2017,Faranda2018}, defined by:
\begin{align}
    G_L^{SM}(r)
    &=
    \begin{cases}
        \mathcal{N}_L e^{-\frac{1}{1-(r/L)^2}} & r \leq L\\
        0 & r > L
    \end{cases}.
    \label{eq:Glr}
\end{align}
where $\mathcal{N}_L$ is a normalisation coefficient. Function (\ref{eq:Glr}) is everywhere smooth, and may be thought of as a Gaussian compactified by bringing the point at infinity to $L$. This kernel is used for all analysis in this paper. Previous results suggest that the local energy budget depends little on the precise choice of filter, for example in experimental \citep{kuzzay2017} and numerical \citep{Johnson2020} experiments for homogeneous isotropic turbulence. 

To compute the $\mathcal{D}_L(\mathbf{u})$ field used in this study, we first calculate velocity increments from the $(u,\upsilon,w)$ field. Then, an angular average (around a circle of great circle radius $r$) is applied at each point. This is the quantity in square brackets in (\ref{eq:int_piphi}). The angular average is taken in the local spherical cap centred at $\mathbf{x}_1$. The calculation of the increments is made by assuming a tangent quadratic structure. There is negligible error in the tropical regions compared to full spherical geometry, and at displacements of $\sim$40 degrees, the error in distances is $\sim$10\%. Our method is also less sensitive to noise because gradients are computed analytically on smooth filter kernels rather than on discrete model fields as is done in other methods (e.g. \citet{aluie2018JPO}). The result of the angular integral is then integrated over great circle separation distances $r$, weighted by the gradient of the filter kernel. Calculations were performed using the open-source software package LoSSETT.

\begin{figure}[t]
 \noindent\includegraphics[width=39pc]{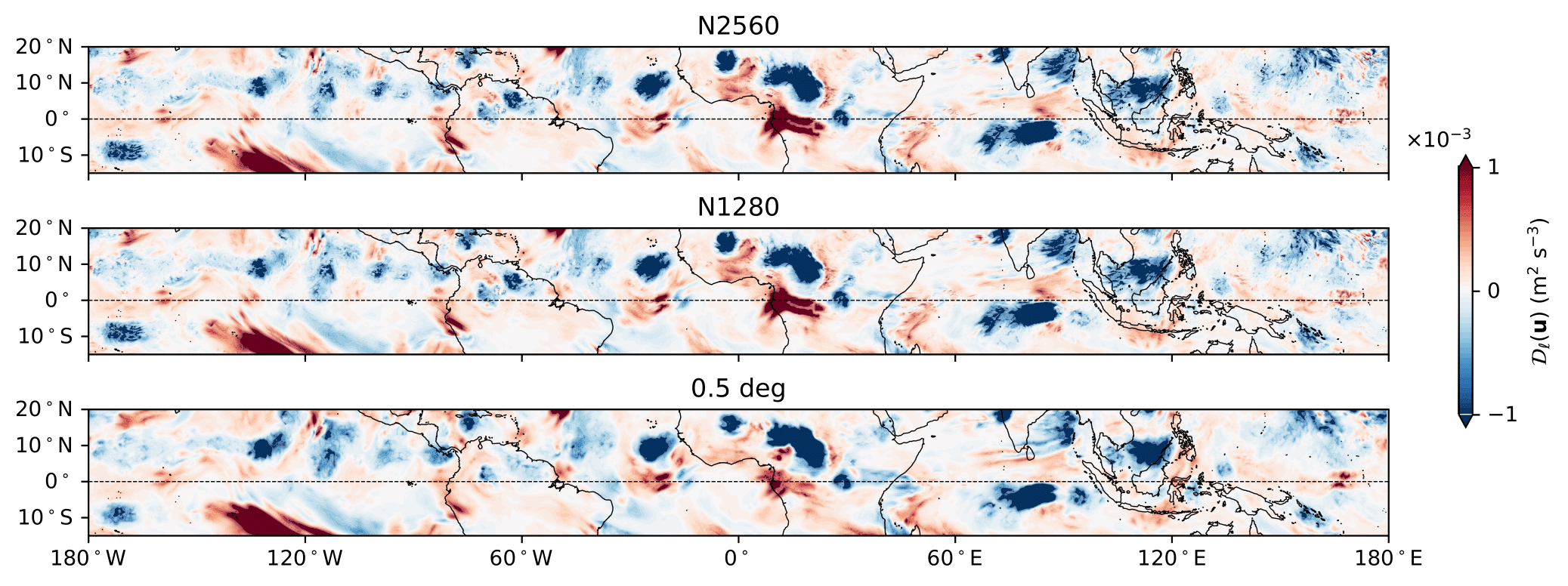}\\
 \caption{Snapshot of interscale energy transfer across a length scale of 440 km at 200hPa for the tropics 15S-20N from the RAL3 N2560 model field on 2016-08-02T0000Z. The three panels represent three different resolutions used in this study: N2560 is the native resolution of the model data; N1280 represents an interpolation to halve the horizontal resolution and quarter the computational expense; 0.5 deg represents the coarsest data used in the study.}\label{fig:inst_mapsnap}
\end{figure}

To give an idea of what the interscale energy transfer field looks like in the tropical atmosphere, an instantaneous snapshot of $\mathcal{D}_L(\mathbf{u})$ from RAL3 for $L = 440$ km at 200 hPa is shown in Fig. \ref{fig:inst_mapsnap}. The large mesoscale of 440 km is used for the majority of figures in this paper, and the energy transfer across this length scale is calculated from N2560 ($\sim$ 5 km latitude $\times$ 8 km longitude at the equator) model data interpolated onto a N1280 ($\sim$10$\times$16 km at the equator) or 0.5 degree ($\sim$ 55 km) grid (depending on length scale) to reduce computational expense. Figure \ref{fig:inst_mapsnap} shows that this choice is justified, since at such a large length scale, the calculation whether made on the original N2560 grid, a grid half as fine or the 0.5 degree grid makes little difference. The 440 km length scale is chosen to facilitate comparison with \citet{Faranda2018}, who elected to study energy transfers across 220 km. The reason our chosen $L$ is twice that of \citet{Faranda2018}'s is that the kernel we use is the same as theirs, but we find it more informative to talk about length scales on the diameter of the kernel rather than the radius. Furthermore, this length scale is well separated from small scale convection and the wavelength of CCEWs. The domain in Fig. \ref{fig:inst_mapsnap} (15S to 20N) is chosen to highlight the activity north of the equator during DYAMOND Summer (boreal summer - the ITCZ is centred north of the equator during this season).

Distinct regions of upscale energy transfer (blue) may be clearly seen in Fig. \ref{fig:inst_mapsnap}, and identified with features such as mesoscale convective systems over Africa, and clusters of convection in the southeastern Indian Ocean where there is an active negative Indian Ocean Dipole (IOD) event. The signal of the convectively coupled Kelvin wave analysed later can be seen as a patch of upscale over the ocean stretching from the coast of peninsular Malaysia to the Philippines. Fig. \ref{fig:inst_mapsnap} compared to Fig. \ref{fig:Dlu_tm} demonstrates the extremely variable-in-time nature of this diagnostic, reflecting the high variability seen in dynamical and convective fields in the tropical atmosphere. The magnitudes of the local and instantaneous values of \DR  reported are on the same order as atmospheric turbulent dissipation rates calculated from observations \citep{Nowak2025}.

\section{Methods and Data}\label{s:m&d}

\subsection{The K-Scale simulations}
The CTC km-scale model domain is being run as part of the UK Met Office's `K-Scale' initiative (\cite{jones2023}, \cite{jones2025II}) which also serves as the UK contribution to the DYAMOND protocol \citep{stevens2019dyamond,takasuka2024DYAMOND3}. The DYAMOND (DYnamics of the Atmosphere Modelled On Non-hydrostatic Domains) protocol is an international collaborative project involving modelling centres across the world exploring CPMs spanning global or large domains at high resolution and their benefits. All centres modelled a common time period. We will focus on the DYAMOND summer period in this study from 2016-08-01 to 2016-09-09. The DYAMOND Summer period was chosen to match up with the NARVAL \citep{NARVAL} observational campaign during these dates. 40 days was chosen to provide a month of data after any spin up and be long enough to separate questions of predictability and internal variability. The relevant tropical climate indices during DYAMOND Summer are listed in Table \ref{tab:DS_indices}. Of particular relevance to this study is that there is no MJO during DYAMOND Summer, meaning CCEWs can be studied as their own system rather than as part of the MJO envelope (e.g. \cite{roundy2008,K09}). The output data used are 3-hourly. The simulations are free-running for 40 days, i.e. with no data assimilation scheme in place like for an operational forecast. This is to allow features to grow by themselves in the model rather than be forced by the initial conditions fed in from the data assimilation scheme.

\begin{table}[t]
\caption{Climate indices during the DYAMOND Summer study period. Abbreviations are the El Ni\~{n}o Southern Oscillation (ENSO), Indian Ocean Dipole (IOD), Madden-Julian Oscillation (MJO), and Quasi-Biennial Oscillation (QBO). The indices were calculated for ENSO using the Nino3.4 index from HadISST data \citep{HadISST1}; the IOD using the Dipole Mode Index from HadISST data; the MJO using the Realtime Multivariate MJO index from NOAA OLR and NCEP/NCAR reanalysis winds data \citep{WheelerHendon04,NCEPNCAR}; the QBO from NCEP/NCAR reanalysis data \citep{NCEPNCAR}; and the BSISO using the all-season OLR-based MJO index (OMI, \citealt{Kiladis2014OMI}).}\label{tab:DS_indices}
\begin{center}
\begin{tabular}{ccccrrcrc}
\hline\hline
  & DYAMOND Summer\\
\hline
Dates & 2016-08-01 to 2016-09-09\\
ENSO & La Ni\~{n}a (Nino3.4r $\sim$ -1$^\circ$C)\\
IOD & Negative decaying (DMI -0.4 $\to$ -0.2$^\circ$C)\\
MJO & None\\
QBO & Westerly ($\sim +10$ m s$^{-1}$)\\
BSISO & Phase 7 $\to$ 8 (weak) and decaying\\
\hline
\end{tabular}
\end{center}
\end{table}

For the analysis carried out in this study, we compare two UM \citep{brown2012UM}  coupled to the Joint UK Land Environment Simulator (JULES, \citealt{best2011JULES,clark2011JULES}) simulations from the K-Scale hierarchy \citep{jones2025II} with differing science configurations. One simulation (referred to as GAL9 N2560 or GAL9 for short throughout) makes use of the Global Atmosphere and Land version 9 (GAL9, \citealt{walters2019GAL}) configuration. The other simulation (RAL3.2 N2560, RAL3 for short) uses the Regional Atmosphere and Land version 3.2 (RAL3.2, \citealt{bush2025RAL3}). Details of the differences between the two simulations are summarised in Table \ref{tab:modelconfigs}.

Both models have identical grids and cover the CTC domain from 40S to 26N. These bounds were chosen to meet the Himalayas in the north and encompass all of southern Africa and the Indian Ocean tropical cyclone basin. The two models are both driven by a global N1280 ($\Delta x \sim 10$ km) model with GAL9 physics. The key difference between GAL9 and RAL3 being investigated in this study is the convection scheme (or lack thereof), with most other features being kept the same.

GAL9 parametrizes deep conveciton. This convection parametrization is not active in RAL3. This means that convection is permitted to explicitly evolve through the model dynamics on the grid. At this resolution there is still a lot of sub-grid convection, and this is handled by the turbulence scheme. The 3D turbulence scheme employed by RAL3 does not force artificial smoothness like the parametrization scheme in GAL, which is how explicit circulations are allowed to develop close to the grid scale. There is additional sub-grid vertical mixing in RAL3 above the boundary layer that is parametrized by a Smagorinsky-Lilly-type eddy diffusivity. Other differences between RAL3 and GAL9 that must be noted are the different cloud fraction schemes: GAL9 uses the PC2 prognostic scheme \citep{wilson2008pc2}, and RAL3 the bimodal diagnostic scheme \citep{KVW2021b,KVW2021bimodal}. The microphysics in GAL9 is parametrized using the \citet{wilsonballard1999} single moment scheme; and RAL3 uses the double moment Cloud AeroSol Interacting Microsphysics scheme (CASIM, \citealt{field2023CASIM}). In the boundary layer, GAL9 employs the 1D turbulent mixing scheme of \citet{lock2000BL}, and RAL3 employs a blended 3D Smagorinsky and 1D scheme, described in \citet{boutle2014}.

\begin{table}[t]
\caption{Summary of differences between the CTC N2560 GAL9 and RAL3 model simulations.}
\label{tab:modelconfigs}
\begin{center}
\begin{tabular}{>{\raggedright\arraybackslash}p{3.5cm} >{\raggedright\arraybackslash}p{5.5cm} >{\raggedright\arraybackslash}p{5.5cm}}
\toprule
\textbf{Model} & \textbf{GAL9} & \textbf{RAL3} \\
\midrule
Convection & Mass flux convection parametrization & No mass flux convection scheme (convection permitting) \\
Large-scale cloud & PC2 - Prognostic & Bimodal - diagnostic\\
Microphysics & Single moment & Double moment CASIM\\
Boundary layer & Unblended 1D & Blended 1D and 3D Smagorinsky
\\
References & \citet{walters2019GAL,tomassini2023,jones2025II} & \citet{bush2025RAL3,jones2025II} \\
\bottomrule
\end{tabular}
\end{center}
\end{table}

The two simulations are across the same domain and for the same time period, and both have the same grid. Both simulations are forced with hourly-updating lateral boundary conditions from the driving model. The lower boundary sea surface temperature is updated daily using 1/20$^\circ$ Operational Sea Surface Temperature and Sea Ice Analysis (OSTIA) data \citep{donlon2012OSTIA}. All simulations are initialised from the UM operational analysis (hereafter referred to as UM analysis, \citealt{clayton2013UMA}), and the analysis is used as a comparative benchmark representing a `perfect forecast' to evaluate the models' performance. The UM analysis is provided at 6-hourly temporal frequency based on the UM Global Atmosphere model GA7 at N1280 resolution \citep{walters2019GAL}. Precipitation observations are from the GPM-IMERG product version 7 \citep{huffman2014GPM}.


\subsection{Equatorial wave filtering}

Equatorial wave dynamical fields are extracted from the model or analysis $(u,\upsilon,Z)$ (zonal wind, meridional wind, geopotential height) fields using the 2D-in-Space Parabolic Cylinder Function (2DS-PCF) method developed by \citet{Y03}. The 2DS-PCF method is applied independently on individual pressure levels, and the vertical structure of the waves can then be inferred by analysis of multiple layers.

Due to the relatively short time period covered by our simulations, the ``realtime" extension of the method \citep{yang2021} is used, whereby a long analysis period is appended to the start of the forecast to enable a time filter with a very broad window so that eastward and westward signals can be distinguished before spatial projection. There are multiple ways to identify equatorial waves from modelled and observed flows, with \citet{Knippertz22} calling for a combination of wavenumber-frequency filtering and spatial projection methods. Since we are concerned here with kinetic energy and the 3D dynamical structure of the waves, the 2DS-PCF method is the most suitable for our purposes.

Equatorial waves are solutions to the shallow water equations (SWEs) on the equatorial beta-plane \citep{Matsuno66}. Substituting zonally propagating wavelike solutions in $y$ (meridional distance) back into the SWEs and rearranging yields a 2nd order ordinary differential equation in $\hat{\upsilon}(y)$ only. Solutions to this 2nd-order ODE are known as the Parabolic Cylinder Functions (PCFs) and are expressed in terms of nondimensionalised $y$:

\begin{equation}
    y^*=y\sqrt{\beta/c}, \quad \hat{\upsilon} = D_n(y^*).
\end{equation}

Here, the constant of separation $c$ (and therefore equivalent depth) is determined by a best fit to reanalysis data \citep{Y03,y07p1,yang2021}. The meridional structure (i.e. structure in $y^*$) of the waves is then given by these PCFs for varying $n$:
\begin{equation}
    D_n(y^*) = \exp\left(-{\frac{y^*}{2}}\right)H_n\left(\frac{y^*}{\sqrt{2}}\right),
\end{equation}
where $H_n$ is an Hermite polynomial of degree $n$ \citep{AB65}. Following \citet{GC74}, in anticipation of the form the expanded equations will take, it is convenient to introduce, in place of $u$ and $Z$, the orthogonal variables $q$ and $r$:

\begin{equation}
	\begin{split}
		q=gZ/c+u,\\
		r=gZ/c-u.
	\end{split}
    \label{eq:qrv}
\end{equation}

Substituting $q$ and $r$ into the SWEs and rearranging allows us to express the waves as a series solution in terms of the basis functions
\begin{equation}
    \begin{split}
        &q = q_0D_0 + q_1D_1 + \sum_{n=1}^{n=\infty} q_{n+1}(x,t)D_{n+1}(y) \\
        &\upsilon = \hspace{13mm} \upsilon_0D_0 + \sum_{n=1}^{n=\infty} \upsilon_n(x,t)D_n(y) \\
        &r = \hspace{26.5mm} \sum_{n=1}^{n=\infty} r_{n-1}(x,t)D_{n-1}(y).
    \end{split}
    \label{eq:PCFs3line}
\end{equation}

Expressing the functions for $q,\upsilon,r$ like this relates the 2D spatial structure of the waves to the dispersion relations by allowing us to express the spatial projection for all $n$ and include the $n=-1$ Kelvin, and $n=0$ westward mixed Rossby-gravity (WMRG) and eastward inertia-gravity (EIG) wave solutions without having to make them a `special consideration' \citep{WheelerNguyen2015}. The way of expressing the spatial structures of the waves in equation (\ref{eq:PCFs3line}) demonstrates how the degree of the polynomial $n$ (the meridional mode) relates to how many wave types can be described by a certain $n$.

The waves are extracted from $(u,\upsilon,Z)$ by the following process. First, $(u,\upsilon,Z)$ is transformed to $(q,r,\upsilon)$ as described by equation (\ref{eq:qrv}). Then, the $(q,r,\upsilon)$ data are subject to a zonal Fast Fourier Transform (FFT) from physical to wavenumber space. Then, a frequency filter (which allows a wide range of timescales from 2-30 days) is applied to partition eastward and westward propagating waves. This is where the most common wave extraction technique \citep{hayashi1971,WK99} stops and categorises the equatorial waves based purely on wavenumber and frequency information. The Hayashi/Wheeler-Kiladis method is applied to data that has been averaged in latitude, assuming either symmetric or anti-symmetric structure about the equator. In the 2DS-PCF method, the meridional structure as well as the zonal wavenumber is used to identify the wave amplitude and phase by spatial projection of the full data onto the orthogonal wave basis, given by equation (\ref{eq:PCFs3line}). The projected data is then transformed back to $(u,\upsilon,Z)$ to obtain the EW dataset. Equatorial wave phase speeds from this data set were then calculated using the Radon transform (See Appendix A for details).




\section{Results and Discussion}\label{s:results&discuss}

\subsection{Simulated waves and precipitation}\label{ss:results1}
\begin{figure}[t]
 \noindent\includegraphics[width=39pc]{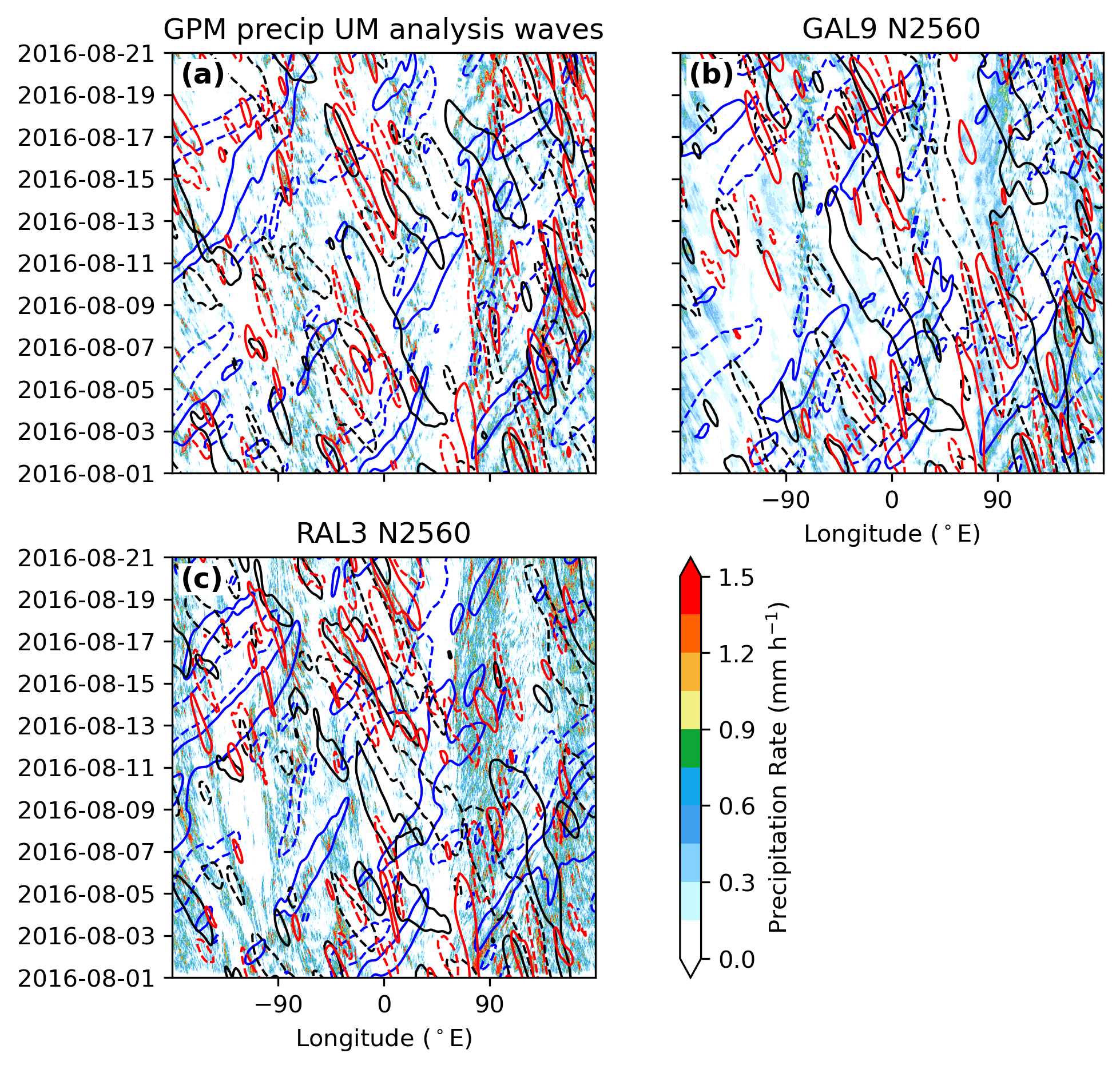}\\
 \caption{Hovm\"{o}ller (longitude-time) diagrams of precipitation (shading) and equatorial wave winds at 850hPa - Kelvin (blue contours), $n=1$ equatorial Rossby (R1) (black contours), WMRG (red contours). Quantities averaged between 10S-10N: precipitation in mm h$^{-1}$, $u_{KW}$ at $\pm$ 1 m s$^{-1}$, $u_{R1}$ and $\upsilon_{WMRG}$ at $\pm$ 2 m s$^{-1}$. Dashed contour lines indicate negative values. GPM precipitation and UM analysis waves in \textbf{(a)} (the observational data set); GAL9 N2560 in \textbf{(b)}, and RAL3 N2560 in \textbf{(c)}. Quantities shown for the first 20 days of the DYAMOND Summer simulation.}\label{fig:precipwaves}
\end{figure}

Hovm\"{o}ller (longitude-time) diagrams of quantities averaged across 10S-10N are shown in Fig. \ref{fig:precipwaves} - the first 20 days are chosen for clarity in the figure, and to allow us to analyse the evolution of simulated features from the same initial conditions. This diagram gives us a first look at the key differences between the RAL3 and GAL9 models and the observations. In terms of precipitation: the GAL9 model has many more stationary features than the RAL3 or observations (e.g. see the features around 90W at the start of the simulation), and its eastward propagating features are quite weak. This key difference between explicit and parametrized convection in simulating propagating precipitation is also found in the study of \citet{dias2018}. The abundance of propagating precipitation in the RAL3 model is more consistent with the observations than the GAL, but appears to simulate too much rain (see for example in the Indian Ocean region around 90E). These features of the models compared to the observations are also reflected in Table \ref{tab:precipfigs}, where GAL9 simulates a higher mean precipitation compared to GPM, albeit with a smaller variance. RAL3 simulates similar variance to the observations, but with a mean that is higher still. All differences between the datasets are significantly different at the 99\% confidence level due to their large size. These differences in precipitation mean and variances between the km-scale models and observations are consistent with the findings of \citet{JRB21}, where they analysed the DYAMOND Summer runs from the MPAS-A model.

All wave-dynamical quantities are computed on a common 0.5$^\circ$ grid, with precipitation on a 0.1$^\circ$ grid. It is worth noting here that although the GPM-IMERG blended observational product made from rain gauge measurements and satellite observations is the best estimate for large-scale precipitation in the tropics, it is still beset by significant biases \citep{DaSilvaGPM}. This is due to a number of factors, including inaccessible rain gauge data on much of the land in the Maritime Continent and equatorial Africa, and lack of observations over the tropical ocean.

\begin{table}[t]
\caption{Mean and variance in precipitation for the observations (GPM) and models shown in Fig. \ref{fig:precipwaves}.}\label{tab:precipfigs}
\begin{center}
\begin{tabular}{ccccrrcrc}
\hline\hline
 & Mean (mm h$^{-1}$) & $\sigma^2$ (mm$^2$ h$^{-2}$)\\
\hline
\textbf{GPM} & 0.17 & 0.08 \\
\textbf{GAL} & 0.19 & 0.04 \\
\textbf{RAL} & 0.20 & 0.08 \\
\hline
\end{tabular}
\end{center}
\end{table}

Where the wave contours travel with the precipitation, this is a clear signal of a Convectively Coupled Equatorial Wave (CCEW). The Convectively Coupled Kelvin Wave (CCKW) initialised on 2nd August at 90E and propagating across the Maritime Continent is present in both models in Figs. \ref{fig:precipwaves} (b) and (c) as well as in the analysis/observations in Fig. \ref{fig:precipwaves} (a). In the observations, this wave propagates eastward until it encounters a convectively coupled Rossby (R1) wave and WMRG over the Maritime Continent at about 120E, resulting in a large amount of precipitation being deposited over the region, and the slowing of the CCKW. This interaction does not occur in the same way in the models in (b) and (c), with there being much weaker R1 and WMRG waves in the models; in RAL3 the KW slows much like in the observations, but the deposition of large amounts of rain is difficult to pick out amongst the very abundant rain in that model, and in the GAL9 the CCKW is terminated altogether. This interaction between the Kelvin, R1 and WMRG waves does not occur in the same way in the two models because the simulated propagation speeds of the waves in RAL3 and GAL9 are different (see Table \ref{tab:phase_speeds_table}). In GAL, the CCKW propagates a similar distance making it as far as the dateline in the central Pacific by Aug 11th, where it also encounters a Rossby wave, albeit in the opposite phase compared to the observations. In the GAL9 model at this point there is also not a strong WMRG like there is in the analysis. In the RAL3, the CCKW propagates farther, continuing after interaction with the easterly-phase R1 wave and terminating 180 degrees and 2 weeks later at 90W, presumably encountering the Andes. This difference may however be a factor of the free-running simulations having diverged from the analysis by 10 days out.

\begin{table}[t]
\caption{Dominant propagation speeds for the three EW types in the UM analysis, and the two models for the data shown in Fig. \ref{fig:precipwaves} calculated using the Radon transform method over the full DYAMOND Summer simulation period. All figures are in m s$^{-1}$. Negative phase speed indicates westward propagation. Variability ranges (1$\sigma$) in brackets.}\label{tab:phase_speeds_table}
\begin{center}
\begin{tabular}{ccccrrcrc}
\hline\hline
 & Kelvin & R1 & WMRG\\
\hline
\textbf{UM analysis} & 12 (2.8) & -9 (3.7) & -10 (3.6)\\
\textbf{GAL} & 15 (10) & -9 (4.3) & -16 (12)\\
\textbf{RAL} & 16 (4) & -15 (9) & -16 (6)\\
\hline
\end{tabular}
\end{center}
\end{table}

The dominant phase speeds of the three main EW types for the UM analysis and the two models are shown in Table \ref{tab:phase_speeds_table}. The observed phase speeds from the analysis were cross-referenced with the extremely comprehensive analyses of \citet{Y07p2,yang2009} where they used 15 years of reanalysis data and satellite-observed convective fields, and are comparable. The models simulate a faster Kelvin wave, but the variability ranges overlap. This is the case for all the models and waves. It is also interesting to note that the models always simulate a higher standard deviation, particularly the GAL. This could be because of a reduced degree of convective coupling in GAL, allowing for a wider range of `dry' (uncoupled with convection) propagating waves. Convective coupling reduces phase speed in CCEWs \citep{K09}. For the R1 wave, the GAL9 simulates the same dominant phase speed with a higher variability, and the RAL3 a much faster phase speed with a much larger variability. The reason for this is because the distribution of phase speeds across the record calculated from the Radon transform \textit{(not shown)} is bimodal. In the analysis these peaks are relatively close together at around 9 and 13 m s$^{-1}$. In the RAL3, the faster peak is strongest at the 15 m s$^{-1}$ value reported in the table, and the second, weaker peak, is at around 6 m s$^{-1}$, hence the large variability range.

\begin{figure}[t]
 \noindent\includegraphics[width=39pc]{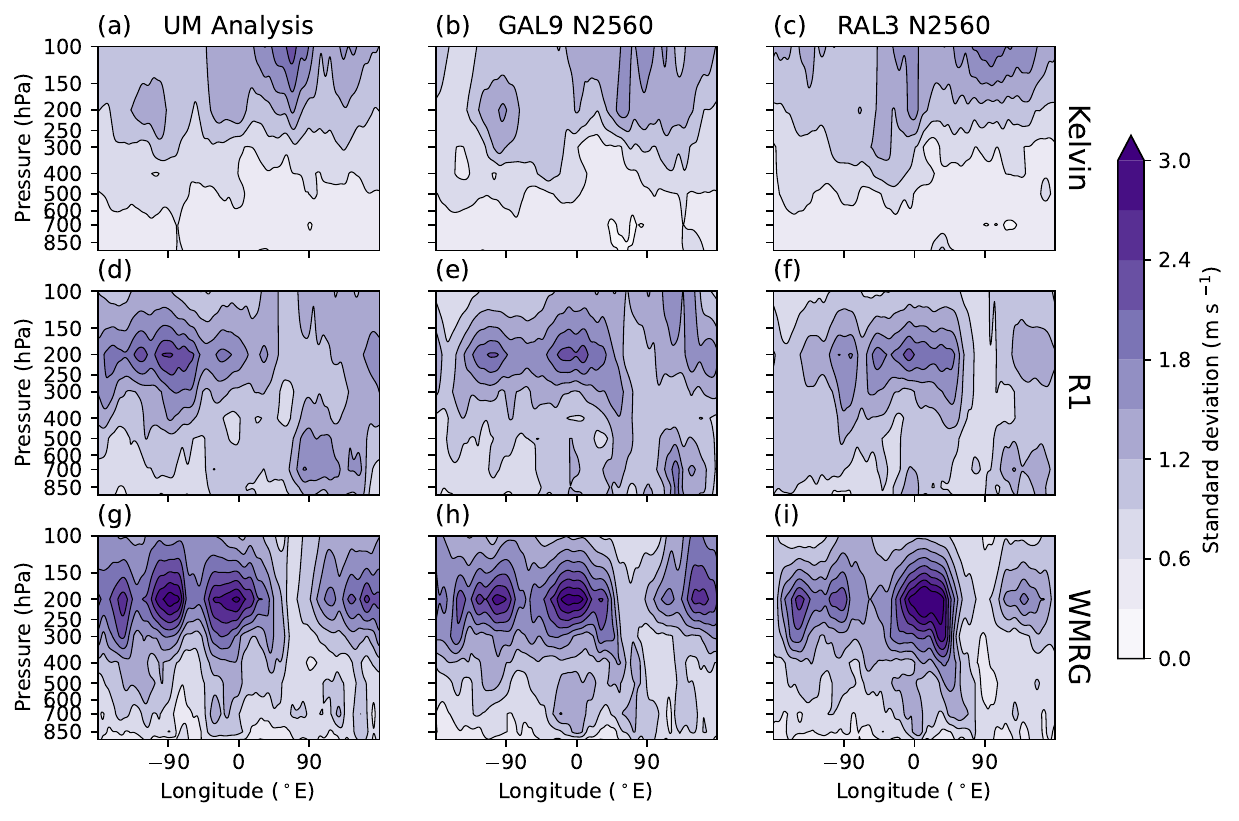}\\
 \caption{Longitude-height slice of wave standard deviation in time for Kelvin (a) - (c), R1 (d) - (f), and WMRG (g) - (i) waves for UM analysis (left column), GAL9 N2560 (middle column), and RAL3 N2560 (right hand column). The standard deviations are calculated on 10S-10N equatorially-averaged Kelvin wave $u$, R1 $u$ and WMRG $\upsilon$.}\label{fig:SD}
\end{figure}

The differences in the simulation of the 3 key wave types can be illustrated by a longitude-height plot of standard deviation in time of wave velocity fields as shown in Fig. \ref{fig:SD}. The six-week average structures in the UM analysis presented here are similar to the 15-year average structures presented in \citet{y07p1,yang2009}. As previously reported by these earlier studies, we expect to see the strongest wave activity in the upper-troposphere for the westward waves and around the tropopause for the Kelvin waves. This is due to the highest wave amplitude being at these levels.  The peak in Kelvin wave variability around 90E at upper-tropospheric levels in the analysis is not fully simulated by either model. It is slightly weaker and further east in RAL3 \textbf{(c)}, and not well defined in GAL9 \textbf{(b)}. There is another peak west of 90W around 200 hPa - this is likely associated with Kelvin waves propagating across the Pacific and terminating at the Andes. This peak is stronger than analysis in GAL, and not present in RAL. The reason for this being stronger in GAL9 and not present in RAL3 could be due to the presence of the gravity wave drag (GWD) parametrization in the GAL9 simulation. This parametrization simulates the subgrid response to the flow blocking by unresolved orography. The N2560 resolution is in the `grey zone' of the orographic GWD (some orography will be resolved, but only a fraction), meaning it is possible that the GWD is simulating too much drag in GAL, and without the parametrization in RAL3, not enough is simulated. Further analysis of this particular parametrization is beyond the scope of this study; see \citet{Vosper2016} for details.

The R1 waves in the analysis in Fig. \ref{fig:SD} \textbf{(d)} exhibit activity centred around 200 hPa in the western hemisphere (WH) - a pattern also seen in reanalysis in \citet{yang2009}. The GAL9 model simulation \textbf{(e)} has a less pronounced peak with wave activity spread over a wider longitudinal range, while the RAL3 \textbf{(f)} has an activity centre that is also weaker but more concentrated, in this case around 0$^\circ$. The middle and lower troposphere structure of the R1 wave in the analysis consists of a peak in activity from 90E and eastwards. For the WMRG waves in Fig. \ref{fig:SD} \textbf{(g)}, there are peaks in activity centred around 200 hPa mostly in the WH (also seen in the analysis product in \cite{yang2009}). GAL9 \textbf{(h)} represents this quite similarly, whereas RAL3 \textbf{(i)} simulates a very strong peak in activity just east of 0$^\circ$, and the peaks in the WH are weaker than in the analysis and GAL. The differences between GAL9 and RAL3 may just be due to noise given the short record analysed, but importantly both are able to represent the general behaviour of variability in time throughout the tropical atmosphere, shown by their similarity to the UM analysis which in turn is qualitatively similar to the patterns displayed in \citet{yang2009}.

From this first analysis of the waves and precipitation in the RAL3 and GAL9 models, we have observed that both models simulate more rain than the observations, the RAL3 particularly so. Excessive tropical rain is commonly observed in CPMs (e.g. \citealt{howard2024}). The GAL9 model does not capture the high variability of precipitation seen in the tropics, whereas the RAL3 does. Again, this has been previously observed by other CPM studies (e.g. \citealt{JRB21}). Regarding equatorial wave propagation, the waves appear comparable on a Hovm\"{o}ller plot but a closer investigation reveals there are differences in simulated propagation speed between the analysis and models, with both models simulating a too fast Kelvin and WMRG wave, and GAL9 with a higher variability of simulated wave speeds. The Rossby wave speeds in GAL9 are comparable to the analysis, and faster in RAL3 with higher variability. The distribution of wave variability in the study period across the equatorial band and full depth of the troposphere is comparable between all models and the analysis.

\subsection{Simulated upscale energy transfer and dynamics}
\begin{figure}[t]
 \noindent\includegraphics[width=39pc]{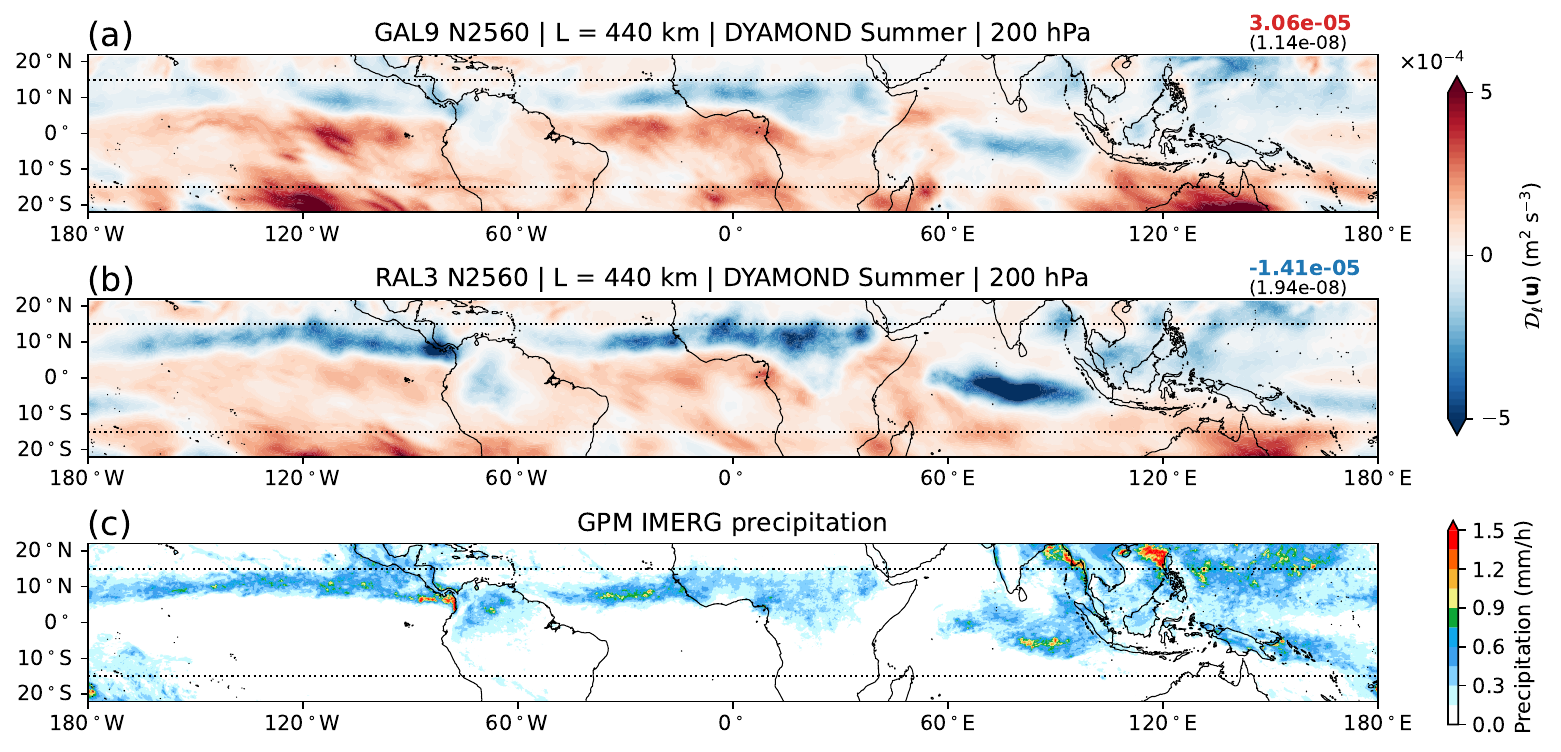}\\
 \caption{Maps of time-mean $\mathcal{D}_L(\mathbf{u})$ for $L$ = 440 km averaged over the DYAMOND Summer simulation period for \textbf{(a)} GAL9 N2560 and \textbf{(b)} RAL3 N2560. Blue represents upscale energy transfer, and red downscale. The maps are for the full tropical channel from 20S to 20N. The average from 15S to 15N (indicated by the dotted lines) is on the top right of each panel, colour-coded to indicate whether the mean transfer in that region is upscale or downscale. Variance in $\mathcal{D}_L(\mathbf{u})$ over the same region indicated by the value in brackets. \textbf{(c)} mean GPM-IMERG precipitation for the same time period.}\label{fig:Dlu_tm}
\end{figure}

To get a first indication of the reasons for the differences in simulated propagating convection between the models, time mean maps of interscale energy transfer for $L$ = 440 km at 200 hPa (the top of the tropical troposphere) are shown in Fig. \ref{fig:Dlu_tm} for GAL9 and RAL3. This is the level where horizontal divergence is at a maximum, and so we expect to see high values of upscale energy transfer, because of the mechanism explained at the end of this subsection. For the interscale energy transfer analysed in this study, we do not compare the models to the analysis product. This is because it is at a much coarser resolution than the models and is driven by a data assimilation (DA) scheme with parametrized convection. This means we would expect to see low magnitudes of interscale energy transfer in the anlaysis that we would not trust at the lower length scales analysed since this would be too close to the grid scale of the analysis. Furthermore, the presence of DA in the analysis product would not allow for the investigation of propagating features developing on their own for a longer time period without the impact of the DA scheme. 

In Fig. \ref{fig:Dlu_tm}, the blue colours indicate the energy is being transferred upscale from scales below 440 km to scales larger than this. The red colours indicate that energy is being transferred downscale from scales larger than 440 km to scales smaller than this. One clear and obvious difference between the two is that in RAL3, there is orders of magnitude more upscale energy transfer, and this is reflected by the deep-tropical mean at 200 hPa being net upscale. The mean was taken over 15S to 15N to avoid impact of sub-tropical systems, such as the subtropical jet. The domain 20S-20N is shown to allow for visualisation of the full extent of the ITCZ in Boreal Summer. The values and spatial patterns shown here are consistent with \citet{kouhen2024} who use a different interscale energy transfer diagnostic (the same as in \cite{aluie2018JPO}) and a different data set (operational ECMWF analysis for the year 2020). 

The ITCZ is visible in Fig. \ref{fig:Dlu_tm} as the blue band of upscale centred around 10N. Similarly, the SPCZ is visible extending west from New Guinea to the other side of the dateline. Both convergence zones are associated with much stronger upscale energy transfer in the RAL3 model. Another key feature in Fig. \ref{fig:Dlu_tm} is the abundant upscale energy transfer in the Indian Ocean. This period coincides with a strong negative IOD event (cf. Table \ref{tab:DS_indices}), which is associated with enhanced convection in the southeast Indian Ocean \citep{saji1999} -- a region which is convectively active even without the negative IOD. Once more, this region of upscale energy transfer is orders of magnitude greater in RAL3 than GAL, saturating the colour scale. Another upscale feature which is much larger and more extensive in RAL3 is the deep convection over the Amazon. Panel \textbf{(c)} in Fig. \ref{fig:Dlu_tm} provides further evidence that the upscale energy transfer at 200hPa is located at the tropical convergence zones. These zones are characterised by consistently high precipitation, and are reflected in the map of mean rainfall over the study period in this panel. The precipitation fields are from GPM-IMERG, and we can see from Fig. \ref{fig:Dlu_tm} that the RAL3 model is simulating a strong ITCZ in terms of 200hPa upscale energy transfer.

\begin{figure}[t]
 \noindent\includegraphics[width=39pc]{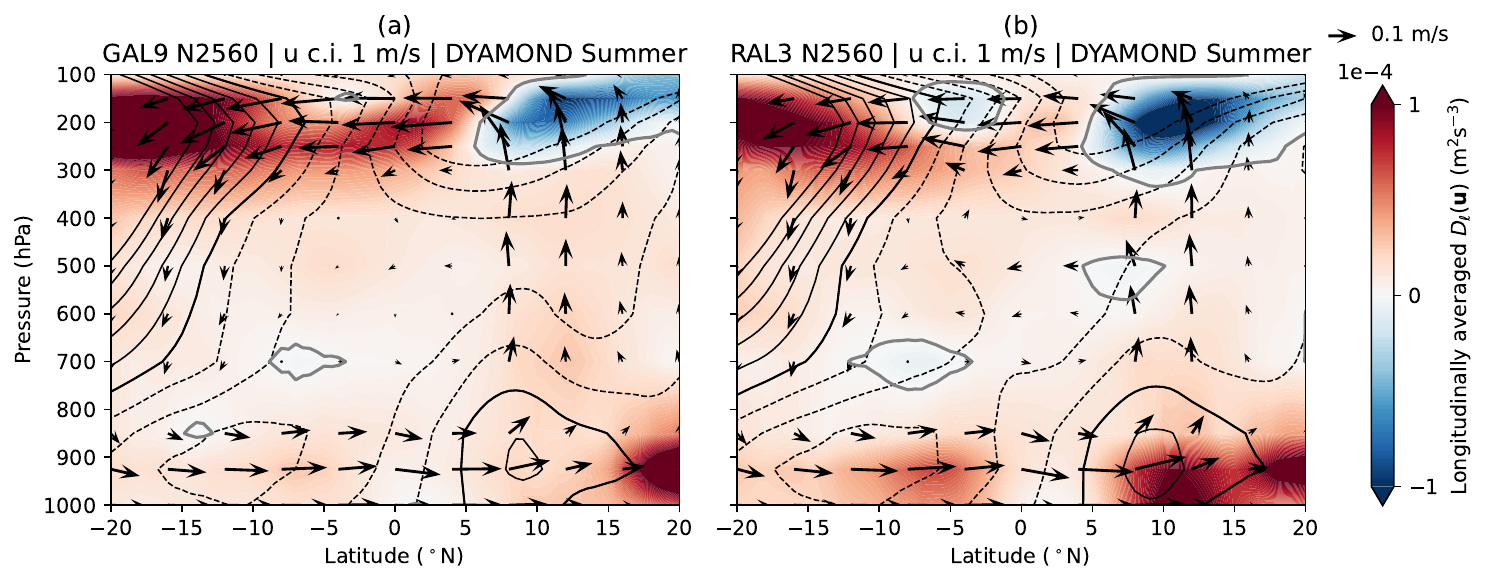}\\
 \caption{Longitudinally averaged latitude-pressure plots of time-mean $\mathcal{D}_L(\mathbf{u})$ for $L$ = 440 km (shading, zero contour thick grey line) for the domain shown in Fig. \ref{fig:Dlu_tm}. Zonal mean zonal wind in black contours in intervals of 1 m s$^{-1}$, and zonal mean meridional and vertical velocities in vectors, with the reference arrow of 0.1 m s$^{-1}$ shown in the top right of the figure. \textbf{(a)}: for the GAL9 N2560 model; \textbf{(b)}: for RAL3 N2560. $w$ has been re-scaled by the aspect ratio of the domain for better visualisation of the time-mean flow.}
 \label{fig:latslice}
\end{figure}

Fig. \ref{fig:latslice} shows longitudinally-averaged, time-mean interscale energy transfer (shading, zero contour thick grey line), zonal mean zonal winds (black contour lines), and the mean meridional circulation (vectors). The main features of Fig. \ref{fig:latslice} which connect back to Fig. \ref{fig:Dlu_tm} are the sections of upscale energy transfer at the top of the troposphere centred around 200 hPa. This area of upscale is associated with time-average flow and divergence in the tropical upper troposphere as a result of deep convection and associated vertical velocities in the ITCZ centred around 10N. Near the surface, downscale transfer is seen associated with convergence as part of the ITCZ circulation. The RAL3 exhibits both stronger upscale and downscale energy transfer associated with the ascending branch of this circulation, suggesting that explicit representation of convection leads to RAL3 simulating interscale energy transfer that is significantly larger than that in the GAL9 model. The large scale Hadley circulation is of similar strength in both RAL3 and GAL, but RAL3 shows an enhanced secondary circulation at 500 hPa associated with convection. Centred around 5S and 150 hPa is a region of upscale energy transfer which can be identified with the region of strong upscale energy transfer over the Indian Ocean seen in Fig. \ref{fig:Dlu_tm}. This region of strong upscale is also associated with downscale energy transfer at low levels underneath the deep convection associated with net flow convergence. As also seen in Fig. \ref{fig:Dlu_tm}, this region is about twice as large in RAL3 compared to GAL. This is further evidence of the CPM's ability to simulate strong upscale energy transfer associated with deep convection. At the southern edge of \textbf{(a)} in the upper troposphere, the signal of the subtropical jet is seen associated with strong zonal winds and strong downscale energy transfer.

In the middle troposphere, there are two areas of weak upscale energy transfer in the RAL3, but only one in the GAL. After consulting maps of time-mean interscale energy transfer at 500 and 600 hPa (not shown), the feature at this level in the RAL3 around 5-10N is associated with the stronger secondary ITCZ circulation in this model, and the presence of cumulus congestus associated with the West African Monsoon. This upscale feature is not present in the GAL. At around 700 hPa between 5 and 10S is another upscale feature which appears in both models. On consultation of the maps at this level this appears to be collocated with upscale energy transfer over the Amazon and Congo rainforests, possibly associated with shallow convection in these regions. 

Interscale energy transfer will be a new diagnostic to many, and we present an heuristic explanation here. Fig. \ref{fig:latslice} shows the archetypal pattern of low-level convergence and associated downscale energy transfer beneath ascent and upper-level divergence with upscale energy transfer. This pattern has been observed in idealised Rayleigh-B\'{e}nard convection studies \citep{Togni2015,Green2020}, and also appears to apply in deep tropical convective systems. It should be noted that although time-mean divergence and upscale energy transfer are correlated, that does not in turn mean that the upscale energy transfer results from time-mean divergent flow --- the interscale energy transfer is not the resolved divergent flux of energy. Instead, the instantaneous divergent outflow acts on air parcels, expanding them and transferring their energy from below or near the scale of interest (e.g. the flow of the updraft as indicated in the schematic) to a larger scale, therefore transferring the parcel's energy upscale. The exact opposite is true in the convergent part of the cell in the lower troposphere, where parcels that are on the scale of a downdraft (similar scale to that of the anvil at the top of the convective plume) are then compressed by the convergent flow down to the scale of the updraft.

The results presented here exhibit differences compared with \citet{Faranda2018} for a number of reasons. Although the method is comparable, visible differences can be seen. We think these differences are most likely because of our use of a 2D kernel rather than the 3D kernel (with undocumented vertical extent and vertical boundary handling) employed by \cite{Faranda2018}. Appendix~B compares the \cite{Faranda2018} method and our method side-by-side for the same instant and the same dataset.

\begin{figure}[t]
 \noindent\includegraphics[width=39pc]{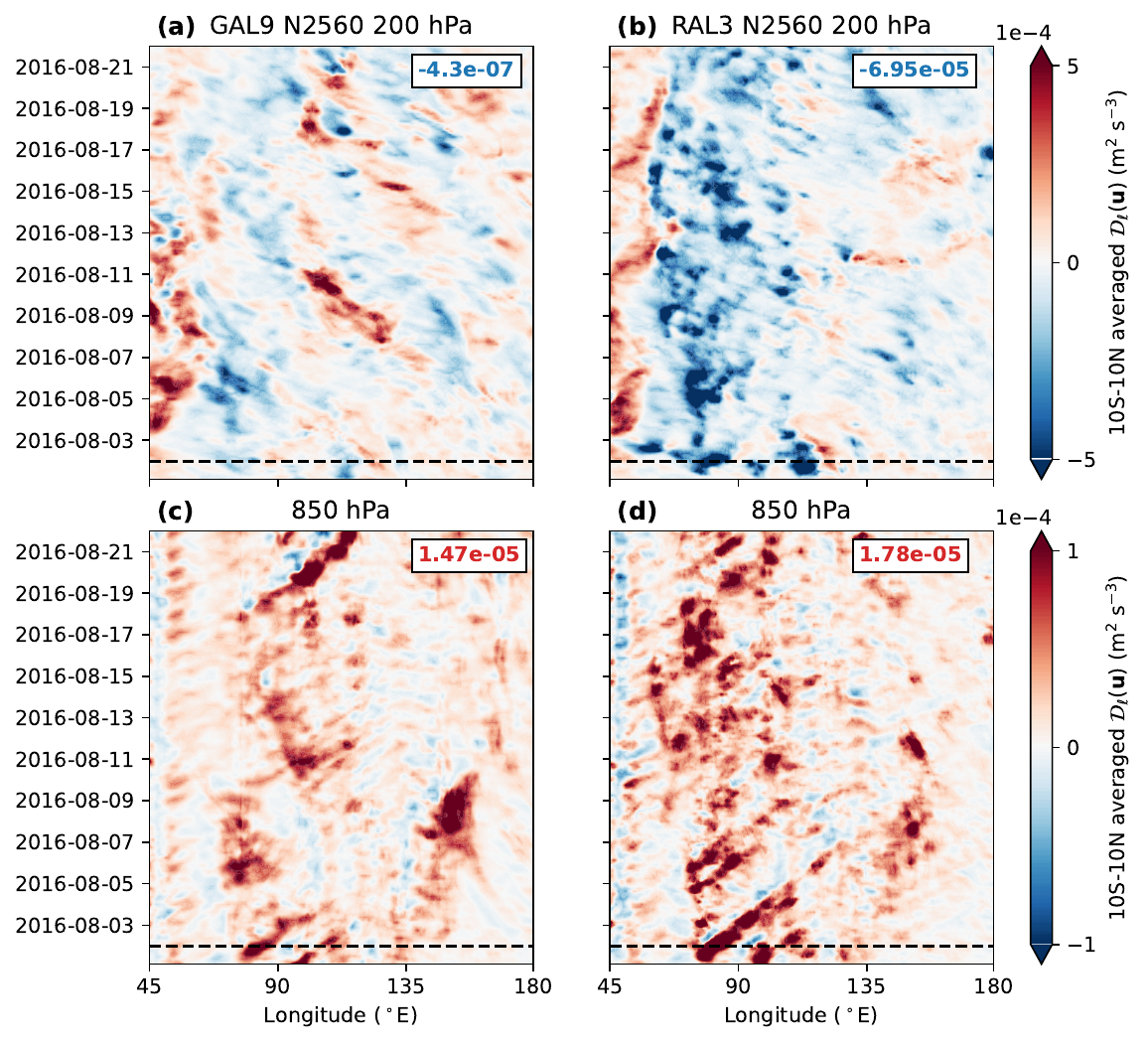}\\
 \caption{Time-longitude (Hovm\"{o}ller) diagrams of $\mathcal{D}_L(\mathbf{u})$ for $L$ = 440 km averaged between 10S and 10N for the 20 day study period, limited to 45 - 180E to remove the strong orographic effects of the East African Highlands at 850 hPa. \textbf{(a)} for GAL9 at 200hPa; \textbf{(b)} RAL3 at 200 hPa; \textbf{(c)} GAL9 at 850 hPa, and \textbf{(d)} RAL3 at 850 hPa. Note the contour limits are 5 times higher in \textbf{(a)} and \textbf{(b)} than in \textbf{(c)} and \textbf{(d)}. Colour-coded numbers in the top right of each panel are mean values of the data shown with the colour indicating net upscale (blue) or net downscale (red). The dashed line indicates the time of the snapshot profile shown in Fig. \ref{fig:Dlu_KW_snapshot}. The time mean has been removed to clarify the propagating features, and this has been done for Fig. \ref{fig:Dlu_KW_snapshot}, and for calculating the composites shown in Fig. \ref{fig:COMPOSITE}.}
 \label{fig:Dlu_hov}
\end{figure}

Moving away from the time-mean picture, a Hovm\"{o}ller of interscale energy transfer is shown in Fig. \ref{fig:Dlu_hov}. The contour limits are 5 times higher for the 200 hPa panels than the 850 hPa panels due to the larger overall amount of kinetic energy due to higher wind speeds at this altitude. For the region shown in Fig. \ref{fig:Dlu_hov} spanning the Indian Ocean, Maritime Continent, and West Pacific, there are abundant eastward-propagating downscale energy transfer features at 850 hPa which are broadly collocated with the propagating rainfall and Kelvin waves in Fig. \ref{fig:precipwaves}. These downscale features are occasionally coincident with upscale energy transfer at 200 hPa (e.g. see in the RAL3 starting at 90E on 2016-08-01), indicative of propagating deep convection (likely a CCKW). The picture at 200 hPa is mostly dominated by westward-propagating features. These Hovm\"{o}ller diagrams were analysed with waves overlain and for symmetric, anti-symmetric, and off-equatorial interscale energy transfer (not shown). We found that the westward-propagating features were indeed indicative of CCEW activity: R1 waves in the case of the off-equatorial interscale energy transfer (0-20N average), and WMRG waves for the anti-symmetric part (calculated by taking the average of the difference between the fields in the 20 degrees north and south of the equator).

\clearpage
\clearpage

\begin{figure}[t]
 \noindent\includegraphics[width=39pc]{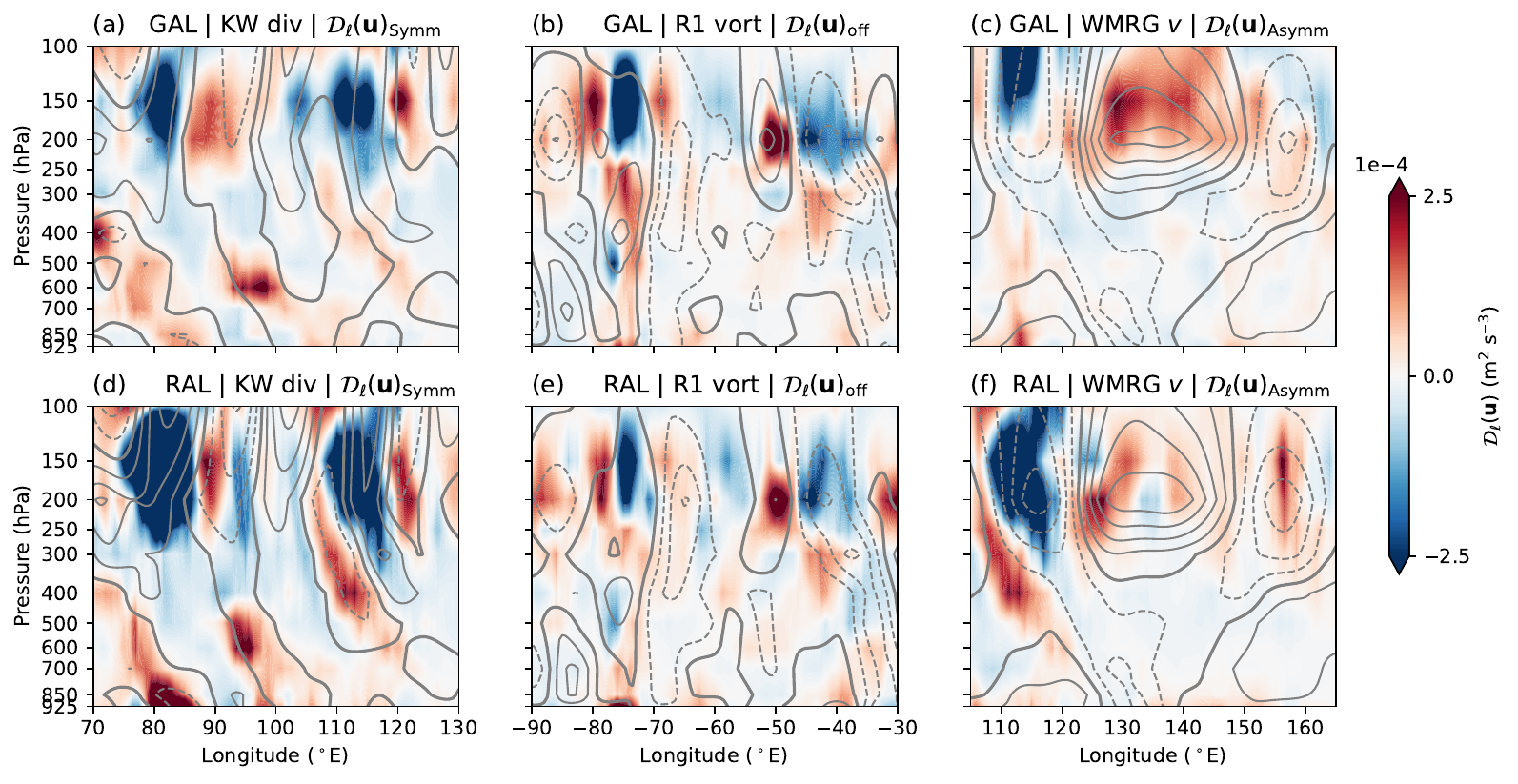}\\
 \caption{\textbf{(a)}: Vertical profile snapshot on 2016-08-02 at 0 UTC of symmetric (10S-10N average) $\mathcal{D}_L(\mathbf{u})$ and KW divergence averaged over the same latitude band in GAL. Contour interval $1\times10^{-4}$ s$^{-1}$. Zero contour heavy line in all panels. \textbf{(d)} as in \textbf{(a)} but for RAL. \textbf{(b)}: off-equatorial (0-20N average) $\mathcal{D}_L(\mathbf{u})$ and R1 vorticity off-equatorial average. Contour interval $5\times10^{-5}$ s$^{-1}$, \textbf{(e)} as in \textbf{(b)} but for RAL. \textbf{(c)}: anti-symmetric ([0$\to$20N - 20S$\to$0]/2) $\mathcal{D}_L(\mathbf{u})$ and WMRG $\upsilon$ averaged over the equatorial band. Contour interval 2 m s$^{-1}$. \textbf{(f)} as in \textbf{(c)} but for RAL.}
 \label{fig:Dlu_KW_snapshot}
\end{figure}

\subsection{Equatorial waves and upscale energy transfer}
The picture of the imprint of the general circulation on the interscale energy transfer can be seen in Fig. \ref{fig:Dlu_hov} by the fact that the 200 hPa regions are net upscale and the 850 hPa regions net downscale. Once again it is evident that RAL3 is simulating significantly more upscale than GAL9 at 200 hPa --- 2 orders of magnitude more in this case. To investigate how propagating features in Fig. \ref{fig:Dlu_hov} are associated with different equatorial waves, longitude-pressure snapshots at the time indicated by the dashed line in Fig. \ref{fig:Dlu_hov} of dynamical wave fields and interscale energy transfer are shown in Fig. \ref{fig:Dlu_KW_snapshot}. All wave snapshots are shown across the longitudes where they are strongest in this season. For the KW this is over the Indian Ocean, for the R1 wave it is over the Atlantic, and for the WMRG over the West Pacific. 

From figures \ref{fig:Dlu_KW_snapshot} \textbf{(a)} and \textbf{(d)}, we can see upscale energy transfer related to Kelvin wave divergence. Upscale energy transfer in the upper troposphere in the divergent phase of the wave is observed because the convecting plume coupled to the wave is expanding near the tropopause, transferring energy from scales smaller than 440 km (the scale of the updraft) to scales larger than 440 km (toward the scale of the wave). This is balanced by downscale energy transfer in the convergent phase of the wave, both beneath the divergence in the lower troposphere, and laterally in the wave troughs either side of the upper-level wave divergence. Another thing to note from Fig. \ref{fig:Dlu_KW_snapshot} is that the vertical structure of the Kelvin wave in RAL3 is closer to observations and theory than GAL9 (cf. \citealt{SK03,y07p1}). The wave is higher amplitude, exhibits clear westward tilt with height, and upward propagation into the stratosphere in RAL3, whereas the GAL9 does not exhibit these qualities as clearly, especially in the lower troposphere.

Turning our attention now to the R1 wave in figures \ref{fig:Dlu_KW_snapshot} \textbf{(b)} and \textbf{(e)}, the dynamical structure is similar in both models. This is the case for both the westward waves. As previously noted by \citet{yang2021}, westward-propagating CCEWs are not as poorly simulated as Kelvin waves in conventional NWP models employing convection parametrization schemes. For the dynamical R1 wave, the structure is as expected for boreal summer \citep{y07p1}, with slight westward tilt and a hint of baroclinic structure. The R1 wave negative (cyclonic) vorticity is collocated with off-equatorial upscale energy transfer. In a convectively coupled R1 wave, the cyclonic phase of the wave has been observed to coincide with deep convection \citep{y07p1,ferrett2020}.

The WMRG snapshot in Fig. \ref{fig:Dlu_KW_snapshot} \textbf{(c)} and \textbf{(f)}, much like the R1, shows similarities between the two model configurations, and a structure expected by theory and observations \citep{y07p1} with less westward tilt with height, and a baroclinic structure. The WMRG wave convergence appears to be collocated with anti-symmetric upscale energy transfer. This is typically expected, since deep convection coupled to dynamical WMRG waves is anti-symmetric (e.g. \cite{WK99,K09}).

\begin{figure}[t]
 \noindent\includegraphics[width=39pc]{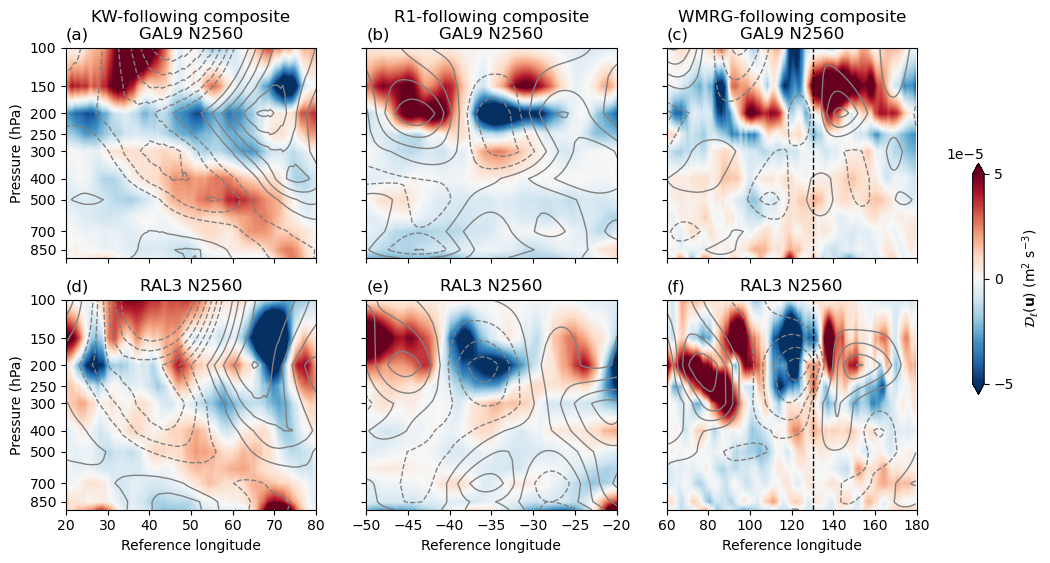}\\
 \caption{As in Fig. \ref{fig:Dlu_KW_snapshot} but for wave-following composites over 10 days as described in the text. ``Reference" longitude is now used to reflect the fact that the values on the figure are the longitudes at time 0, which shift in time with the composite before averaging. Shading contour limits are five times lower than in in Fig. \ref{fig:Dlu_KW_snapshot} to account for averaging along the wave-following axis. The Kelvin divergence and WMRG $\upsilon$ contour intervals are half that in Fig. \ref{fig:Dlu_KW_snapshot}, with R1 vorticity contour intervals being quartered. The vertical dashed lines on (c) and (f) indicate the regions over which the pattern correlations and covariances were computed for the results shown in Fig. \ref{fig:CORRLINES}.}
 \label{fig:COMPOSITE}
\end{figure}

The purpose of Fig. \ref{fig:Dlu_KW_snapshot} was to connect the propagating interscale energy transfer features seen in Fig. \ref{fig:Dlu_hov} with equatorial waves. However at a single time, it is not possible to capture the propagation direction of this quantity. To examine the association of interscale energy transfers with propagating wave features, wave-following composites are presented in Fig. \ref{fig:COMPOSITE}. Wave-following composites at the speed of each modelled wave as shown in Table \ref{tab:phase_speeds_table} were calculated as follows:

\begin{enumerate}
    \item Remove the time mean from the interscale energy transfer field
    \item Take the symmetric, off-equatorial or anti-symmetric mean
    \item Shift the relevant quantity (interscale energy transfer here) following the wave speed of interest at each time step for the length of time desired. In this case we choose 10 days to allow us to analyse the full lifetime of one wave.
    \item Average along the tilted axis to obtain a longitude-pressure slice of wave-composited interscale energy transfer.
\end{enumerate}

\begin{figure}[t]
 \noindent\includegraphics[width=39pc]{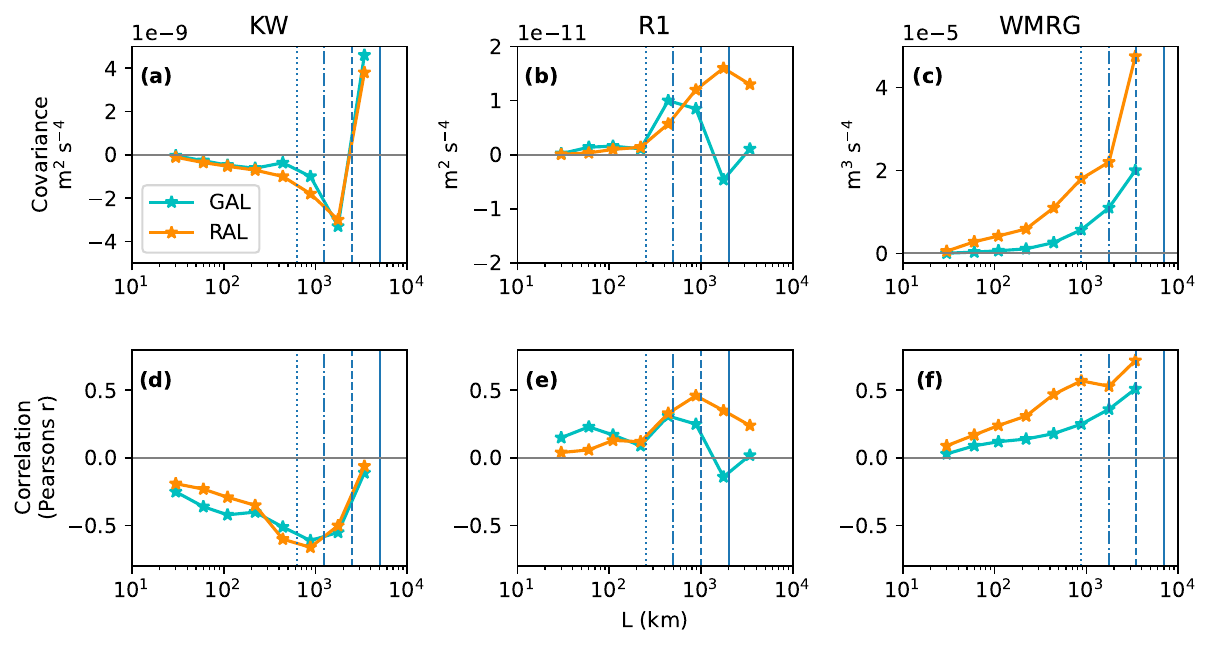}\\
 \caption{Covariances and correlations between the fields shown in Fig. \ref{fig:COMPOSITE} as a function of characteristic length scale $L$ . Kelvin wave divergence about the equator is compared with equatorially symmetric KW-following interscale energy transfer, R1 off-equatorial vorticity with off-equatorial R1-following interscale energy transfer, and WMRG cross-equatorial $\upsilon$ with anti-symmetric  WMRG-following interscale energy transfer, as in Fig. \ref{fig:COMPOSITE}. Correlations and covariances for GAL9 in cyan, for RAL3 in orange. The blue vertical lines represent multiples of the composite dominant wavelength, with the dotted line $\lambda/8$, dot-dash $\lambda/4$, dashed $\lambda/2$, and solid line at the dominant wavelength, $\lambda$.}
 \label{fig:CORRLINES}
\end{figure}

The composites in Fig. \ref{fig:COMPOSITE} provide more information than the instantaneous structures seen in Fig. \ref{fig:Dlu_KW_snapshot}, demonstrating the systematic features that are present over time in interscale energy transfer and dynamical equatorial wave fields. Regarding the Kelvin wave and symmetric eastward-propagating interscale energy transfer, it can be seen that the dynamical wave composite is similar in both the RAL3 and the GAL9 models. As expected, there is more extreme magnitudes of upscale energy transfer in the RAL3, and this appears to be more highly correlated with the KW divergence in the upper troposphere than the GAL. The correlations between the propagating interscale energy transfer and corresponding equatorial waves for varying $L$ are outlined in Fig. \ref{fig:CORRLINES}.

We must make the point here that although there is an obvious correlation between the upscale energy transfer and the wave divergence in the KW composites, it does not in turn mean that the upscale transfer results from the divergent flow - the interscale energy transfer is not the resolved divergent flux of energy. A plausible reason for the association seen here is that deep convection and associated convective-scale outflow is much more likely where there is wave divergence at upper levels (and convergence beneath, in the lower troposphere). The divergent outflow itself expands air parcels, transferring their energy upscale, whereas the opposite is true in the presence of a convergent flow.

Figures \ref{fig:COMPOSITE} \textbf{(b)} and \textbf{(e)} show again the positive correlation between R1 vorticity and interscale energy transfer. This positive correlation between off-equatorial interscale energy transfer and R1 vorticity is reflected in Fig. \ref{fig:CORRLINES}, with correlations in RAL3 being on average 10\% higher than in GAL, and more consistent (same sign for all $L$).

The WMRG composites span a greater longitudinal extent than the other waves in Fig. \ref{fig:COMPOSITE} \textbf{(c)} and \textbf{(f)}. This is to allow room to show the striking differences in their simulation between the models. In the RAL3 there is the characteristic vertical propagation of the WMRG wave typically seen during the westerly phase of the QBO \citep{yang2012QBO}. The QBO is indeed in a westerly phase during DYAMOND Summer, see Table \ref{tab:DS_indices}. No such vertical propagation is seen in the GAL. The GAL9 does however have a slightly stronger baroclinic wave structure, even if the wave is much weaker. The composite is again consistent with the snapshot for RAL3, with southward WMRG winds being collocated with anti-symmetric upscale energy transfer. This collocation is also present in GAL9 albeit with lower values of interscale energy transfer.

Covariance and correlations between symmetric eastward, off-equatorial westward, and anti-symmetric westward interscale energy transfer and the dynamical waves are presented for varying length scales in Fig. \ref{fig:CORRLINES}. The correlations and covariances displayed here are comparing the dynamical wave and interscale energy transfer composites shown in Fig. \ref{fig:COMPOSITE} for multiple length scales. For reference, the fifth data points for each line represent the $L=440$ km length scale plotted in Fig. \ref{fig:COMPOSITE}.

For the Kelvin wave in Fig. \ref{fig:CORRLINES} (a) and (d), we can see that the covariance between the dynamical wave and interscale energy transfer is stronger in the RAL3 for length scales 880 km and lower, comparable and at a maximum for $L=1760$ km, and then the other side of the zero line for the maximum length scale analysed here of 3410 km. Consultation of the composites at these two large length scales (not shown) reveals that the reason for this is because at $L=1760$ km, the regions of strong interscale energy transfer are close to the scale of the wave (between $\lambda/4$ and $\lambda/2$) and almost in-phase with them. This means a correlation of over 0.5 in both models, and with a highly negative covariance - negative (upscale) collocated with positive (divergent) wave-dynamical flow. However at the 3410 km length scale, the interscale energy transfer features are double the size, and are almost large enough to stretch across an entire wavelength with the same sign, leading to correlations close to 0 and a positive covariance, since the deepest regions of upscale energy transfer are located in the convergent phase of the wave. Correlations are strongest for $L = 880$ km. The reasons for this will be discussed at the end of this section in conjunction with the information from the other waves. Correlations are slightly larger in magnitude in the RAL3 for 440 and 880km, and comparable for both models at 1760 km.

For the R1 wave in Fig. \ref{fig:CORRLINES} (b) and (e), there are positive covariances for all length scales, except at 1760 km where the GAL9 model negative covariance. This occurs  because a length scale of 1760 km is almost exactly at the scale of the composite R1's wavelength (around 2000 km). In RAL3, the interscale energy transfer at this length scale is large and in phase with the wave, whereas in GAL9 it is out of phase, hence the negative covariance and correlation. Correlations at length scales of 220 km and below are small and positive in both models. Correlations between the R1 composite vorticity structure and the interscale energy transfer are highest for $L = 880$ km in RAL3, and at $L = 440$ km in GAL9. This represents close to a half and a quarter of the wavelength, respectively.

In the WMRG wave in Fig. \ref{fig:CORRLINES} (c) and (f), covariances between WMRG meridional wind and interscale energy transfer are consistently positive for all length scales, and higher in RAL3 compared to GAL. The wavelength of the WMRG wave in this study period is large (around 7000 km), so the maximum interscale energy transfer length scale considered is at $\lambda/2$. The relationship is the same for the correlations, with a positive trend as the length scale of interscale energy transfer increases towards the scale of the wave. Analysing larger length scales would not make sense for this study since the kernel is on a spherical cap, and for large synoptic or planetary scales, information from subtropical and even midlatitude motions would be included, meaning the interscale energy transfer values would not be meaningfully connected to the equatorial wave of interest. This is acknowledged as a limitation of this study, but could be ameliorated if a search was conducted for lower wavelength WMRG waves in future work.

For the Kelvin and R1 waves, where the length scales of interscale energy transfer considered begin to approach the scale (wavelength) of the wave composites, the correlations are largest in the range $[\lambda/8, \lambda/4]$. At these particular length scales, upscale transfer (deep convection) is often in phase with the dynamical wave. Convection coupled to equatorial waves exhibiting length scales in this range has been qualitatively observed (e.g. \citealt{y07p1}), but here we are able to show quantitatively that the strongest convective coupling occurs close to the 880 km length scale. One interpretation of this that requires further investigation is that energy is transferred straight from the large mesoscale of organised convection at 880 km to the wave scale at about 4000 km.

The other interesting comparison we can make here is that this relationship between convection on scales $\lambda/8$ to $\lambda/4$ and the dynamical wave at scale $\lambda$ is redolent of a fundamental property of inverse cascades in 2D turbulence. The idealised study of \citep{chen2006} that states that ``...energy flux across a length $L$ in the inverse [upscale] cascade range is due to thinning of subscale vortices 4-8 times smaller than $L$ by strain at length scale $L$...". \citet{storer2023} also found that energy transferred upscale across a mesoscale in the ocean was principally deposited at a scale 3.6 times larger. To our knowledge, this study represents the first instance of a similar result in the atmosphere. This relationship occurs more coherently in RAL3 than in GAL, with higher correlations and higher covariances between the interscale energy transfer and the wave at these length scales.

\section{Conclusions}\label{s:concs}

In this paper, equatorial waves and their associated local interscale energy transfers have been assessed in two km-scale cyclic tropical channel models where the only major difference is that one has a mass flux convection parametrization scheme and the other model is convection permitting. This paper presents the first calculation of local interscale energy transfers in a km-scale model of the atmosphere. We also present the first analysis of local interscale energy transfers in the equatorial atmosphere, and their connection to large-scale dynamics. We have presented the first finding that the tropical atmosphere in a km-scale model exhibits a key property of inverse cascades in 2D turbulence: energy transferred upscale across a scale $L$ is most strongly correlated with scales approximately two to eight times larger.

The general conclusions are that the RAL3 (convection permitting) and GAL9 (convection parametrized) models simulate similar dynamical structures of equatorial waves, albeit with differences in vertical structures and vertical propagation. RAL3 simulates 56\% more upscale energy transfer than GAL9 in the troposphere, with similar magnitudes of downscale transfer. We hypothesise that these differences in energy transfer are because RAL3 explicitly represents horizontal motions associated with deep convection, whereas the parametrization scheme in GAL9 acts to remove vertical instability in a column without the associated convective circulation.

RAL3 also exhibits higher correlations between this transfer and the dynamical wave, which we hypothesise is as a result of propagating convection travelling more coherently with the waves (higher degree of convective coupling). The correlations between interscale energy transfer and the dynamical wave peak at $L = 880$ km in the Kelvin wave, and this correlation is 8\% higher in RAL3 than GAL. The `amplitude' of the trough down to the maximum correlation is also more prominent in RAL3 than in GAL9 by eye. This is accompanied by 80\% higher covariance in RAL3, reflecting the higher amplitudes of upscale energy transfer.

In the R1 wave, interscale energy transfer is most correlated with wave vorticity at $L = 880$ km in RAL3 and at $L = 440$ km in GAL9. The maximum correlation in RAL3 for the R1 wave is 48\% higher than in GAL9. This pattern is repeated in the covariances, with RAL3 maximum covariance between the wave and R1-following interscale energy transfer being 60\% higher than in GAL9.

In the WMRG wave, due to its large wavelength, there is no discernible peak in correlation or covariance between interscale energy transfer and wave meridional flow at the largest scales accessible in these simulations. The correlations and covariances increase with increasing scale, with correlations averaging 85\% higher in RAL3 across all scales, and covariances averaging 171\% higher.

These results have shown that CPMs simulate more scale coupling between the small scales and large-scale dynamics in the tropical atmosphere. We hypothesise that this increased coupling is due to the explicit representation of convection. The conclusions drawn here are however limited in that only one simulation of 40 days for each model was analysed. It has been shown that RAL3 simulates more strongly convectively coupled EWs than GAL9 through the proxy of propagating interscale energy transfer and precipitation. Although the dynamical waves at any given level may appear similar in amplitude and abundance between the models (Fig. \ref{fig:precipwaves}), the propagating convection through either of the proxies used is very different between the two models, with RAL3 exhibiting an average of 47\% more convective coupling. Rigorously demonstrating that interscale energy transfer propagating with an equatorial wave is an effective proxy for convective coupling is beyond the scope of this study, and left to future investigation. It has also been shown that the equatorial waves simulated by the two models are typically propagating too fast, and with a higher degree of variability compared to the UM analysis. This variability is larger in GAL9 than in RAL3 for CCKWs and the WMRG wave, but smaller for R1 waves.

We find that interscale energy transfer propagates along with CCEWs, with the RAL3 model simulating more propagation and deeper, more coherent structures in interscale energy transfer. These propagating features were investigated by the superposition of wave-filtered flows over interscale energy transfer in an instantaneous (Fig. \ref{fig:Dlu_KW_snapshot}) and wave-following composite (Fig. \ref{fig:COMPOSITE}) visualisation. It is found that in the wave-comoving frame, upscale energy transfer is associated with Kelvin wave divergence, R1 cyclonic vorticity, and WMRG southward cross-equatorial meridional velocity.

It has also been shown that in a time-mean sense, the RAL3 model simulates more upscale and less downscale than GAL9 over the tropics, with RAL3 being net upscale over the DYAMOND Summer study period in the upper troposphere, and GAL9 net downscale. RAL3 exhibits higher spatial variance in interscale energy transfer than GAL. This time mean behaviour in local interscale energy transfer is reflected in the large-scale circulation, with upscale being associated with deep convection in the ITCZ, shallow convection over rainforests, and monsoon flow. Downscale associated with the subtropical jet is seen in both models and they exhibit similar time-mean Hadley cell circulation.

The findings presented in this study are significant for informing future use of convection permitting models in weather forecasting. We have clarified that a NWP model employing a traditional mass flux convection parametrization scheme simulates weaker upscale coupling to equatorial waves than a convection permitting model. It is possible that this is one of the reasons behind the comparatively poor simulation of Kelvin waves in these models. This study has also demonstrated that cyclic tropical channel convection-permitting models are a useful tool to study and understand the multi-scale coupled tropical atmosphere.

A key aim of this paper was to make the connection between simulated interscale energy transfer and equatorial waves in km-scale models. Reflecting this is the key finding that the simulated equatorial waves in the two models studied are different because there is different wave-associated upscale energy transfer and different convection, therefore different degrees of convective coupling. Future work should investigate the implications this has for weather forecasting, and whether an increase in simulated upscale energy transfer leads to robust improvements in forecasts of CCEW-associated precipitation. There is also ample opportunity for future investigation into the fundamental dynamics behind interscale energy transfers in the atmosphere.

%

%

\clearpage
\acknowledgments
EMG acknowledges support from SCENARIO DTP NERC Grant No. NE/5007261/1 and Met Office CASE funding. DS is supported by the UPSCALE project, funded by the UK Department for Science, Innovation \& Technology (DSIT) National Capability AI Programme and managed by the Met Office. The authors wish to thank Dr. G.-Y. Yang (NCAS Reading) for productive discussions on equatorial waves, and Drs. D. Faranda (CNRS-LSCE) and V. Lembo (ISAC-CNR) for useful discussions on interscale energy transfer. We thank two anonymous reviewers for careful comments that improved the clarity and robustness of the paper. The Met Office Unified Model cyclic tropical channel simulations were developed and initialised by the K-Scale team at the Met Office and data shared with us via the JASMIN HPC cluster. Met Office K-Scale research has been supported by the Met Office Weather and Climate Science for Service Partnership (WCSSP) Southeast Asia project under the International Science Partnerships Fund.


%
%
\datastatement Post-processed output from the K-Scale models and UM analysis is available on license from the Met Office via the JASMIN HPC cluster. GPM-IMERG precipitation data version 7 freely available at \url{https://gpm.nasa.gov/}. The code used to compute the interscale energy transfer fields used in this study is freely available at \citet{LoSSETT}; the code to compute the equatorial wave projection at \citet{waves}; all other analysis scripts available from the corresponding author upon request.

%

\appendix[A]




\appendixtitle{Estimating average equatorial wave phase speeds using the Radon transform}

The Radon transform (RT; \cite{Radon1917} -- see \citet{deans2007radon} for up-to-date translation) was initially used for signal processing after its theoretical beginnings \citep{deans2007radon}. The RT was first used on oceanographic data by \citet{challenor2001} to calculate phase speeds of oceanic Rossby waves, and not long after by \citet{Y07p2} to calculate phase speeds of equatorial waves in the atmosphere. The steps of the Radon transform are as follows. See the appendix of \citet{Y07p2} for more details.

\begin{enumerate}
    \item The wavelike field in the longitude-time domain (this can be averaged over several latitudes, or at just one latitude) $f(x,y)$ is projected onto a line at a range of angles $0 < \theta < 180^\circ$ with respect to the $x$ (or longitude) axis.
    \item When the line is perpendicular to the alignment of the crests and troughs of the wave, the projected sum of data from the $(x,y)$ plane to the $x'$ axis, $p(x'_i,\theta)$ has maximum variance; $\sum_i{p^2(x'_i,\theta)}$ is a maximum. See \citet{challenor2001}'s Fig. 1 and \citet{Y07p2}'s Fig. A2 for visualisation.
    \item The angle $\theta$ perpendicular to that projection gives the direction of travel and the phase speed can be derived trigonometrically:
    \begin{equation}
        c_{\mathrm{dominant}} = \tan\left(\theta_{\mathrm{dominant}}\right); \quad \theta_{\mathrm{dominant}} = \mathrm{argmax}_\theta f(\theta)
    \label{eq:cfromRadon}
    \end{equation}
\end{enumerate}

%


\appendix[B]
\appendixtitle{Comparison with Faranda et al. (2018)}

\begin{figure}[t]
    \noindent\includegraphics[width=19pc]{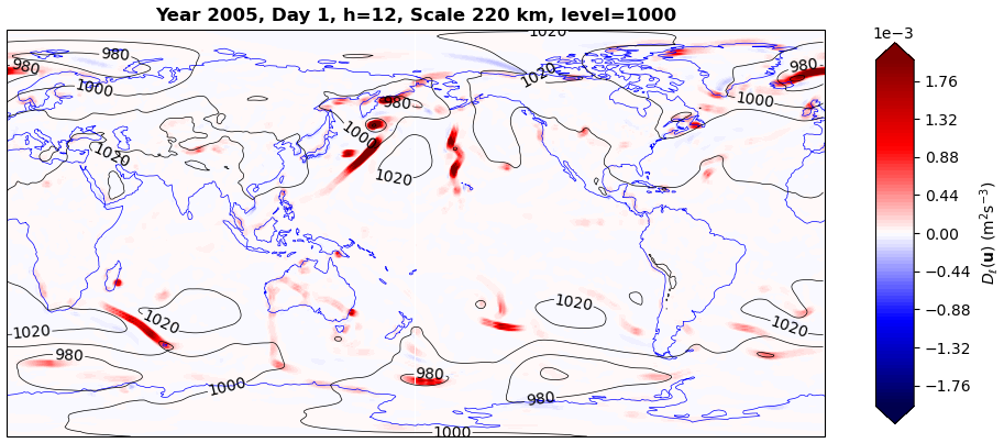}\\
    (a)\\[6pt]
    \includegraphics[width=19pc]{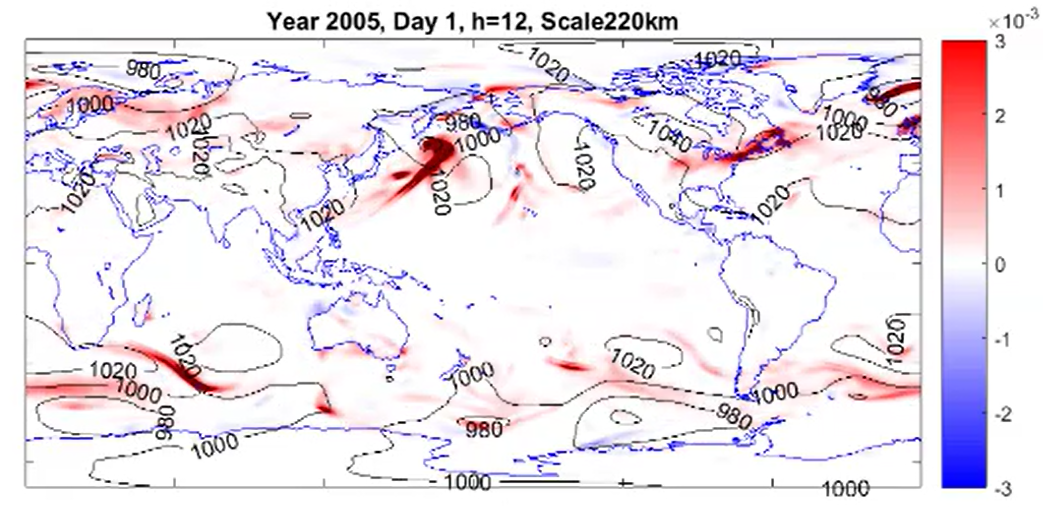}\\
    (b)\\
    \caption{\textbf{(a)}: LoSSETT run on ERA-Interim for 2005-01-01T12:00. \textbf{(b)}: The Faranda et al. (2018) MATLAB programme run on the same instance of data.}\label{fig:F18vLO}
\end{figure}

The method for calculating local inter-scale energy transfers in this paper is based on \cite{Faranda2018} (hereafter F2018); to our knowledge the only other studies to have attempted to calculate maps of local interscale energy transfer in the atmosphere are F2018 and \citet{kouhen2024}. The differences between our approach and that of F2018 are listed here and the consistency between our results are shown. Our approach uses 2D filter kernels (averaging in the horizontal plane), while F2018 use a 3D filter kernel with undocumented vertical extent and vertical boundary handling. Other differences in methodology (for instance, treatment of the Earth's curvature) are expected to be negligible at the small length scales presented in F2018. The data used in F2018 is coarser than the km-scale data analysed in this study. Figure \ref{fig:F18vLO} compares LoSSETT run on ERA-Interim \citep{Dee2011ERAI} with a snapshot from F2018's supplementary material to allow a side-by-side comparison with their method. The same strong downscale features are picked up in the midlatitude storm tracks.

The hypothesis that the 2D vs. 3D kernel explains the majority of differences between our results and those of \cite{Faranda2018} is supported by the fact that Figure 4.2 in \citet{kouhen2024} is consistent with our Figure \ref{fig:Dlu_tm} for a different time period, different dataset, different kernel, and different method of calculating the interscale energy transfer, right down to the units and magnitude. This was calculated using the method outlined in \citet{aluie2018JPO}, which uses a 2D kernel to compute an alternate definition of interscale energy transfer commonly used in spatial filtering frameworks. This leads us to trust the output of our method as well as trust the patterns of time-mean interscale energy transfer we see in the tropical regions. A further difference between our method, F2018's and Kouhen's is the treatment of the curvature of Earth. As mentioned earlier, we expect it to have negligible effects for these small length scales, but for completeness we note here that F2018 operate in the tangent plane, this study on a quadratic curved surface, and the kernel in the method used by \citet{kouhen2024} commutes with derivatives on the sphere.

\bibliographystyle{ametsocV6}
\bibliography{references}

@STRING{AIP 	= "Amer. Inst. Phys."}

@STRING{AN        = "Astrophys.\ Norv."}

@STRING{AR        = "Atmos.\ Res."}

@STRING{MA        = "Meteor.\ Appl."}

@STRING{OCEAN     = "Oceanography"}

@STRING{TELLUS    = "Tellus"}

@article{Matsuno66,
  title={Quasi-Geostrophic Motions in the Equatorial Area},
  author={Taroh Matsuno},
  journal={J. Met. Soc. Japan. Ser. II},
  volume={44},
  pages={25-43},
  year={1966},
  doi={10.2151/jmsj1965.44.1_25}
}

@book{AB65,
title       = {Handbook of mathematical functions},
author      = {Abramowitz, M. and Stegun, I. A.},
year        = {1965},
publisher   = {Dover},
location    = {New York},
pages       = {1043},	
}

@article{dias2018,
  title={Equatorial waves and the skill of NCEP and ECMWF numerical weather prediction systems},
  author={Dias, Juliana and Gehne, Maria and Kiladis, George N and Sakaeda, Naoko and Bechtold, Peter and Haiden, Thomas},
  journal={Monthly Weather Review},
  volume={146},
  number={6},
  pages={1763--1784},
  year={2018},
  doi={10.1175/MWR-D-17-0362.1}
}

@article{takasuka2024DYAMOND3,
  title={A protocol and analysis of year-long simulations of global storm-resolving models and beyond},
  author={Takasuka, Daisuke and Satoh, Masaki and Miyakawa, Tomoki and Kodama, Chihiro and Klocke, Daniel and Stevens, Bjorn and Vidale, Pier Luigi and Terai, Christopher R},
  journal={Progress in Earth and Planetary Science},
  volume={11},
  number={1},
  pages={66},
  year={2024},
  publisher={Springer},
  doi={10.1186/s40645-024-00668-1}
}

@article{GC74,
  title={Wind-induced upwelling, coastal currents and sea-level changes},
  author={Adrian Edmund Gill and Allan J. Clarke},
  journal={Deep Sea Research},
  year={1974},
  volume={21},
  pages={325-345},
  doi={10.1016/0011-7471(74)90038-2}
}

@article {Y03,
      author = "Gui-Ying Yang and Brian Hoskins and Julia Slingo",
      title = "Convectively Coupled Equatorial Waves: A New Methodology for Identifying Wave Structures in Observational Data",
      journal = "Journal of the Atmospheric Sciences",
      year = "2003",
      publisher = "American Meteorological Society",
      address = "Boston MA, USA",
      volume = "60",
      pages=      "1637 - 1654",
      doi = "10.1175/1520-0469(2003)060<1637:CCEWAN>2.0.CO;2"
}

@article{K09,
author = {Kiladis, George N. and Wheeler, Matthew C. and Haertel, Patrick T. and Straub, Katherine H. and Roundy, Paul E.},
title = {Convectively coupled equatorial waves},
journal = {Reviews of Geophysics},
volume = {47},
year = {2009},
doi={10.1029/2008RG000266}
}

@article{judt2018,
  title={Insights into atmospheric predictability through global convection-permitting model simulations},
  author={Judt, Falko},
  journal={Journal of the Atmospheric Sciences},
  volume={75},
  number={5},
  pages={1477--1497},
  year={2018},
  publisher={American Meteorological Society},
  doi={10.1175/JAS-D-17-0343.1}
}

@ARTICLE{NCEPNCAR,
       author = {{Kalnay}, E. and {Kanamitsu}, M. and {Kistler}, R. and {Collins}, W. and {Deaven}, D. and {Gandin}, L. and {Iredell}, M. and {Saha}, S. and {White}, G. and {Woollen}, J. and {Zhu}, Y. and {Leetmaa}, A. and {Reynolds}, B. and {Chelliah}, M. and {Ebisuzaki}, W. and {Higgins}, W. and {Janowiak}, J. and {Mo}, K.~C. and {Ropelewski}, C. and {Wang}, J. and {Jenne}, Roy and {Joseph}, Dennis},
        title = "{The NCEP/NCAR 40-Year Reanalysis Project.}",
      journal = {Bulletin of the American Meteorological Society},
         year = 1996,
        month = mar,
       volume = {77},
       number = {3},
        pages = {437-472},
          doi = {10.1175/1520-0477(1996)077<0437:TNYRP>2.0.CO;2},
       adsurl = {https://ui.adsabs.harvard.edu/abs/1996BAMS...77..437K}
}

@incollection{WheelerNguyen2015,
  author       = {Matthew C. Wheeler and Hanh Nguyen},
  title        = {Tropical Meteorology: Equatorial Waves},
  booktitle    = {Encyclopedia of Atmospheric Sciences},
  editor       = {G. North and F. Zhang and J. Pyle},
  publisher    = {Elsevier},
  year         = {2015},
  month        = {February},
  edition      = {2},
  doi          = {10.1016/B978-0-12-382225-3.00414-X},
  url          = {https://doi.org/10.1016/B978-0-12-382225-3.00414-X}
}

@article {Y07p2,
      author = "Gui-Ying Yang and Brian Hoskins and Julia Slingo",
      title = "Convectively Coupled Equatorial Waves. Part II: Propagation Characteristics",
      journal = "Journal of the Atmospheric Sciences",
      year = "2007",
      publisher = "American Meteorological Society",
      address = "Boston MA, USA",
      volume = "64",
      number = "10",
      pages=      "3424 - 3437",
      doi = "10.1175/JAS4018.1"
}

@article{challenor2001,
  title={Use of the 3D Radon transform to examine the properties of oceanic Rossby waves},
  author={Challenor, Peter G and Cipollini, Paolo and Cromwell, David},
  journal={Journal of Atmospheric and Oceanic Technology},
  volume={18},
  number={9},
  pages={1558--1566},
  year={2001},
  doi={10.1175/1520-0426(2001)018<1558:UOTRTT>2.0.CO;2}
}

@article{stevens2019dyamond,
  title={DYAMOND: the DYnamics of the Atmospheric general circulation Modeled On Non-hydrostatic Domains},
  author={Stevens, Bjorn and Satoh, Masaki and Auger, Ludovic and Biercamp, Joachim and Bretherton, Christopher S and Chen, Xi and D{\"u}ben, Peter and Judt, Falko and Khairoutdinov, Marat and Klocke, Daniel and others},
  journal={Progress in Earth and Planetary Science},
  volume={6},
  pages={1--17},
  year={2019},
  publisher={Springer},
  doi={10.1186/s40645-019-0304-z}
}

@article{HadISST1,
  title={Global analyses of sea surface temperature, sea ice, and night marine air temperature since the late nineteenth century},
  author={Rayner, Na A and Parker, De E and Horton, EB and Folland, Chris K and Alexander, Lisa V and Rowell, DP and Kent, Elizabeth C and Kaplan, A},
  journal={Journal of Geophysical Research: Atmospheres},
  volume={108},
  number={D14},
  year={2003},
  publisher={Wiley Online Library},
  doi={10.1029/2002jd002670}
}

@article{boutle2014,
  title={Seamless stratocumulus simulation across the turbulent gray zone},
  author={Boutle, IA and Eyre, JEJ and Lock, AP},
  journal={Monthly Weather Review},
  volume={142},
  number={4},
  pages={1655--1668},
  year={2014},
  doi={10.1175/MWR-D-13-00229.1}
}

@article{huffman2014GPM,
  title={Integrated multi-satellite retrievals for GPM (IMERG), version 4.4},
  author={Huffman, George and Bolvin, D and Braithwaite, D and Hsu, K and Joyce, R and Xie, P},
  journal={NASA’s precipitation processing center},
  year={2014},
}

@article{howard2024,
  AUTHOR = {Howard, E. and Woolnough, S. and Klingaman, N. and Shipley, D. and Sanchez, C. and Peatman, S. C. and Birch, C. E. and Matthews, A. J.},
TITLE = {Evaluation of multi-season convection-permitting atmosphere -- mixed-layer ocean simulations of the Maritime Continent},
JOURNAL = {Geoscientific Model Development},
VOLUME = {17},
YEAR = {2024},
NUMBER = {9},
PAGES = {3815--3837},
URL = {https://gmd.copernicus.org/articles/17/3815/2024/},
DOI = {10.5194/gmd-17-3815-2024}
}

@article{Kiladis2014OMI,
  title={A comparison of OLR and circulation-based indices for tracking the MJO},
  author={Kiladis, George N and Dias, Juliana and Straub, Katherine H and Wheeler, Matthew C and Tulich, Stefan N and Kikuchi, Kazuyoshi and Weickmann, Klaus M and Ventrice, Michael J},
  journal={Monthly Weather Review},
  volume={142},
  number={5},
  pages={1697--1715},
  year={2014},
  doi={10.1175/MWR-D-13-00301.1}
}

@article{chen2006,
  title={Physical mechanism of the two-dimensional inverse energy cascade},
  author={Chen, Shiyi and Ecke, Robert E and Eyink, Gregory L and Rivera, Michael and Wan, Minping and Xiao, Zuoli},
  journal={Physical review letters},
  volume={96},
  number={8},
  pages={084502},
  year={2006},
  publisher={APS},
  doi={10.1103/PhysRevLett.96.084502}
}

@article{hayashi1971,
  title={A generalized method of resolving disturbances into progressive and retrogressive waves by space Fourier and time cross-spectral analyses},
  author={Hayashi, Yoshikazu},
  journal={Journal of the Meteorological Society of Japan. Ser. II},
  volume={49},
  number={2},
  pages={125--128},
  year={1971},
  publisher={Meteorological Society of Japan}
}

@article{dong2025MCS,
  title={Comparison of Global Mesoscale Convective System Simulations in a Global Storm-Resolving Model and a High-Resolution General Circulation Model},
  author={Dong, Wenhao and Zhao, Ming and Guo, Huan and Harris, Lucas and Cheng, Kai-yuan and Zhou, Linjiong and Ramaswamy, V},
  journal={Journal of Climate},
  volume={38},
  number={10},
  pages={2339--2356},
  year={2025},
  publisher={American Meteorological Society},
  doi={10.1175/JCLI-D-24-0303.1},
}

@article{kolmogorov1941,
  title={The local structure of turbulence in incompressible viscous fluid for very large Reynolds},
  author={Kolmogorov, Andrey Nikolaevich},
  journal={Numbers. In Dokl. Akad. Nauk SSSR},
  volume={30},
  pages={301},
  year={1941}
}

@article{judt2020,
  title={Atmospheric predictability of the tropics, middle latitudes, and polar regions explored through global storm-resolving simulations},
  author={Judt, Falko},
  journal={Journal of the Atmospheric Sciences},
  volume={77},
  number={1},
  pages={257--276},
  year={2020},
  doi={10.1175/JAS-D-19-0116.1}
}

@article{yang2012QBO,
  title={The influence of the QBO on the propagation of equatorial waves into the stratosphere},
  author={Yang, Gui-Ying and Hoskins, Brian and Gray, Lesley},
  journal={Journal of the Atmospheric Sciences},
  volume={69},
  number={10},
  pages={2959--2982},
  year={2012},
  doi={10.1175/JAS-D-11-0342.1}
}

@article{jones2023,
  title={Impact of domain size on tropical precipitation within explicit convection simulations},
  author={Jones, Richard W and Sanchez, Claudio and Lewis, Huw and Warner, James and Webster, Stuart and Macholl, Joshua},
  journal={Geophysical Research Letters},
  volume={50},
  pages={e2023GL104672},
  year={2023},
  publisher={Wiley Online Library},
  doi={10.1029/2023GL104672}
}

@article{jones2025II,
    author = {Jones, Richard W and Lewis, Huw and Sanchez, Claudio and Warner, James},
    title = {Developing a model hierarchy to explore upscale processes in kilometre-scale weather and climate models} ,
    journal = {Bulletin of the American Meteorological Society},
    year = {In review}
}

@article{Nowak2025,
author = {Nowak, Jakub L. and Wacławczyk, Marta and Vassilicos, John. C. and Król, Stanisław and Malinowski, Szymon P.},
title = {Scale-by-scale budget of turbulence kinetic energy in the convective atmospheric boundary layer: Analysis of structure functions},
journal = {Quarterly Journal of the Royal Meteorological Society},
volume = {151},
number = {767},
pages = {e4879},
doi = {10.1002/qj.4879},
year = {2025}
}

@article{Togni2015,
    title={Physical and scale-by-scale analysis of Rayleigh–Bénard convection},
    volume={782}, 
    DOI={10.1017/jfm.2015.547}, 
    journal={Journal of Fluid Mechanics}, 
    author={Togni, Riccardo and Cimarelli, Andrea and De Angelis, Elisabetta}, 
    year={2015}, 
    pages={380–404}}

@article{ferrett2020,
author = {Ferrett, Samantha and Yang, Gui-Ying and Woolnough, Steven J. and Methven, John and Hodges, Kevin and Holloway, Christopher E.},
title = {Linking extreme precipitation in Southeast Asia to equatorial waves},
journal = {Quarterly Journal of the Royal Meteorological Society},
volume = {146},
number = {727},
pages = {665-684},
doi = {10.1002/qj.3699},
year = {2020}
}

@article{lock2000BL,
  title={A new boundary layer mixing scheme. Part I: Scheme description and single-column model tests},
  author={Lock, AP and Brown, AR and Bush, MR and Martin, GM and Smith, RNB},
  journal={Monthly weather review},
  volume={128},
  number={9},
  pages={3187--3199},
  year={2000},
  doi={10.1175/1520-0493(2000)128<3187:ANBLMS>2.0.CO;2}
}

@article{KVW2021b,
  title={A bimodal diagnostic cloud fraction parameterization. Part II: Evaluation and resolution sensitivity},
  author={Van Weverberg, Kwinten and Morcrette, Cyril J and Boutle, Ian},
  journal={Monthly Weather Review},
  volume={149},
  number={3},
  pages={859--878},
  year={2021},
  doi={10.1175/MWR-D-20-0230.1}
}

@article{yang2021,
  title={Real-time identification of equatorial waves and evaluation of waves in global forecasts},
  author={Yang, Gui-Ying and Ferrett, Samantha and Woolnough, Steve and Methven, John and Holloway, Chris},
  journal={Weather and Forecasting},
  volume={36},
  number={1},
  pages={171--193},
  year={2021},
  doi={10.1175/WAF-D-20-0144.1}
}

@article{holloway2013,
  title={The effects of explicit versus parameterized convection on the MJO in a large-domain high-resolution tropical case study. Part I: Characterization of large-scale organization and propagation},
  author={Holloway, Christopher E and Woolnough, Steven J and Lister, Grenville MS},
  journal={Journal of the Atmospheric Sciences},
  volume={70},
  number={5},
  pages={1342--1369},
  year={2013},
  doi={10.1175/JAS-D-14-0308.1}
}

@article{birch2014,
  title={A seamless assessment of the role of convection in the water cycle of the West African Monsoon},
  author={Birch, Cathryn E and Parker, DJ and Marsham, JH and Copsey, D and Garcia-Carreras, L},
  journal={Journal of Geophysical Research: Atmospheres},
  volume={119},
  number={6},
  pages={2890--2912},
  year={2014},
  publisher={Wiley Online Library},
  doi={10.1002/2013JD020887}
}

@article{ferrett2021,
  title={Evaluating convection-permitting ensemble forecasts of precipitation over Southeast Asia},
  author={Ferrett, Samantha and Frame, Thomas H A and Methven, John and Holloway, Christopher E and Webster, Stuart and Stein, Thorwald H M and Cafaro, Carlo},
  journal={Weather and Forecasting},
  volume={36},
  number={4},
  pages={1199--1217},
  year={2021},
  doi={10.1175/WAF-D-20-0216.1}
}

@article{DR2000,
    author = {Duchon, Jean and Robert, Raoul},
    title = {Inertial energy dissipation for weak solutions of the incompressible Euler and Navier-Stokes equations},
    journal = {Nonlinearity},
    year = {2000},
    volume = {13},
    pages = {249-255},
    doi = {10.1088/0951-7715/13/1/312}
}

@article{Leray1934,
    author = {Leray, Jean},
    title = {Essai sur le mouvement d'un liquide visqueux emplissant l'espace},
    journal = {Acta Mathematica},
    year = {1934},
    volume = {63},
    pages = {193-248},
}

@article {Faranda2018,
      author = "Davide Faranda and Valerio Lembo and Manasa Iyer and Denis Kuzzay and Sergio Chibbaro and Francois Daviaud and Berengere Dubrulle",
      title = "Computation and Characterization of Local Subfilter-Scale Energy Transfers in Atmospheric Flows",
      journal = "Journal of the Atmospheric Sciences",
      year = "2018",
      publisher = "American Meteorological Society",
      address = "Boston MA, USA",
      volume = "75",
      number = "7",
      doi = "10.1175/JAS-D-17-0114.1",
      pages=      "2175 - 2186"
}

@article{Green2020,
    author = {Green, Gerritt and Vlyakov, Dimitar G and Mellado, Juan Pedro and Wilczek, Michael},
    title = {Resolved energy budget of superstructure in Rayleigh-Benard convection},
    journal = {Journal of Fluid Mechanics},
    year = {2020},
    volume = {887},
    pages = {A21},
    doi = {10.1017/jfm.2019.1008}
}

@article{kuzzay2015,
  title={Global vs local energy dissipation: The energy cycle of the turbulent von K{\'a}rm{\'a}n flow},
  author={Kuzzay, Denis and Faranda, Davide and Dubrulle, B{\'e}reng{\`e}re},
  journal={Physics of Fluids},
  volume={27},
  number={7},
  year={2015},
  publisher={AIP Publishing},
  doi={10.1063/1.4923750}
}

@article{Augier2012,
    author = {Augier, P and Galtier, S and Billant, P},
    title = {Kolmogorov laws for stratified turbulence},
    journal = {Journal of Fluid Mechanics},
    year = {2012},
    volume = {709},
    pages = {659-670},
    doi = {10.1017/jfm.2012.379}
}

@article{selzcraig2015GRL,
  title={Simulation of upscale error growth with a stochastic convection scheme},
  author={Selz, Tobias and Craig, George C},
  journal={Geophysical Research Letters},
  volume={42},
  number={8},
  pages={3056--3062},
  year={2015},
  publisher={Wiley Online Library},
  doi={10.1002/2015GL063525}
}

@article{Radon1917,
    author = {Radon, J.},
    title = {Über die Bestimmung von Funktionen durch ihre Integralwerte längs gewisser Mannigfaltigkeiten},
    journal = {Math.-Phys. Kl.},
    year = {1917},
    volume = {69},
    pages = {262-267}
}

@article{WK99,
  title={Convectively coupled equatorial waves: Analysis of clouds and temperature in the wavenumber--frequency domain},
  author={Wheeler, Matthew and Kiladis, George N},
  journal={Journal of the Atmospheric Sciences},
  volume={56},
  number={3},
  pages={374--399},
  year={1999},
  doi={10.1175/1520-0469(1999)056<0374:CCEWAO>2.0.CO;2}
}

@phdthesis{kouhen2024,
  author    = {Kouhen, Salah},
  title     = {Local power spectra of the {E}arth's atmosphere},
  school    = {University of Oxford},
  year      = {2024},
  type      = {DPhil thesis},
  url       = {https://ora.ox.ac.uk/objects/uuid:7e7a7bcb-495f-434c-a74e-48b0440d6b03},
}

@article{Knippertz22,
  title={The intricacies of identifying equatorial waves},
  author={Knippertz, Peter and Gehne, Maria and Kiladis, George N and Kikuchi, Kazuyoshi and Rasheeda Satheesh, Athul and Roundy, Paul E and Yang, Gui-Ying and {\v{Z}}agar, Nedjeljka and Dias, Juliana and Fink, Andreas H and others},
  journal={Quarterly Journal of the Royal Meteorological Society},
  volume={148},
  number={747},
  pages={2814--2852},
  year={2022},
  publisher={Wiley Online Library},
  doi={10.1002/qj.4338}
}

@inproceedings{shutts2005kinetic,
  title={Kinetic energy backscatter for NWP models and its calibration},
  author={Shutts, Glenn},
  booktitle={Proc. Workshop on Representation of Subgrid Processes Using Stochastic-dynamic Models},
  pages={13--24},
  year={2005},
}

@inproceedings{ShuttsPalmer2004,
  title={The use of high-resolution numerical simulations of tropical circulation to calibrate stochastic physics schemes},
  author={Shutts, Glenn and Palmer, Tim N},
  booktitle={Proc. ECMWF Workshop Intra-Seasonal Variability},
  pages={83--102},
  year={2004},
  organization={ECMWF Reading, UK}
}

@article{saji1999,
  author       = {Saji, N. H. and Goswami, B. N. and Vinayachandran, P. N. and Yamagata, T.},
  title        = {A dipole mode in the tropical Indian Ocean},
  journal      = {Nature},
  year         = {1999},
  volume       = {401},
  number       = {6751},
  pages        = {360--363},
  doi          = {10.1038/43854}
}

@article{Hill2002,
    author = {Hill, Reginald J},
    title = {Exact second-order structure function relationships},
    journal = {Journal of Fluid Mechanics},
    year = {2002},
    volume = {468},
    pages = {317-326},
    doi = {10.1017/S0022112002001696},
}

@article{cafaro2021,
  title={Do convection-permitting ensembles lead to more skillful short-range probabilistic rainfall forecasts over tropical East Africa?},
  author={Cafaro, Carlo and Woodhams, Beth J and Stein, Thorwald H M and Birch, Cathryn E and Webster, Stuart and Bain, Caroline L and Hartley, Andrew and Clarke, Samantha and Ferrett, Samantha and Hill, Peter},
  journal={Weather and Forecasting},
  volume={36},
  number={2},
  pages={697--716},
  year={2021},
  doi={10.1175/WAF-D-20-0172.1}
}

@article{stein2015,
  title={The representation of the West African monsoon vertical cloud structure in the Met Office Unified Model: An evaluation with CloudSat},
  author={Stein, Thorwald H M and Parker, Douglas J and Hogan, Robin J and Birch, Cathryn E and Holloway, Christopher E and Lister, Grenville MS and Marsham, John H and Woolnough, Steven J},
  journal={Quarterly Journal of the Royal Meteorological Society},
  volume={141},
  number={693},
  pages={3312--3324},
  year={2015},
  publisher={Wiley Online Library},
  doi={10.1002/qj.2614}
}

@article{WKW2000,
  title={Large-scale dynamical fields associated with convectively coupled equatorial waves},
  author={Wheeler, Matthew and Kiladis, George N and Webster, Peter J},
  journal={Journal of the Atmospheric Sciences},
  volume={57},
  number={5},
  pages={613--640},
  year={2000},
  doi={10.1175/1520-0469(2000)057<0613:LSDFAW>2.0.CO;2}
}

@book{deans2007radon,
  title={The Radon transform and some of its applications},
  author={Deans, Stanley R},
  year={2007},
  publisher={Courier Corporation}
}

@article{yang2009,
  title={Convectively coupled equatorial waves in high-resolution Hadley Centre climate models},
  author={Yang, Gui-Ying and Slingo, Julia and Hoskins, Brian},
  journal={Journal of Climate},
  volume={22},
  number={8},
  pages={1897--1919},
  year={2009},
  doi={10.1175/2008JCLI2630.1}
}

@article{ferrett2023,
  title={Hybrid dynamical--statistical forecasts of the risk of rainfall in Southeast Asia dependent on equatorial waves},
  author={Ferrett, Samantha and Methven, John and Woolnough, Steven J and Yang, Gui-Ying and Holloway, Christopher E and Wolf, Gabriel},
  journal={Monthly Weather Review},
  volume={151},
  number={8},
  pages={2139--2152},
  year={2023},
  publisher={American Meteorological Society},
  doi={10.1175/MWR-D-22-0300.1}
}

@article{WheelerHendon04,
  title={An all-season real-time multivariate MJO index: Development of an index for monitoring and prediction},
  author={Wheeler, Matthew C and Hendon, Harry H},
  journal={Monthly weather review},
  volume={132},
  number={8},
  pages={1917--1932},
  year={2004},
  publisher={American Meteorological Society},
  doi={10.1175/1520-0493(2004)132<1917:AARMMI>2.0.CO;2}
}

@article{aluie2018JPO,
  title={Mapping the energy cascade in the North Atlantic Ocean: The coarse-graining approach},
  author={Aluie, Hussein and Hecht, Matthew and Vallis, Geoffrey K},
  journal={Journal of Physical Oceanography},
  volume={48},
  number={2},
  pages={225--244},
  year={2018},
  doi={10.1175/JPO-D-17-0100.1}
}

@article{EyinkAluie2009,
  title={Localness of energy cascade in hydrodynamic turbulence. I. Smooth coarse graining},
  author={Eyink, Gregory L and Aluie, Hussein},
  journal={Physics of Fluids},
  volume={21},
  number={11},
  year={2009},
  publisher={AIP Publishing},
  doi={10.1063/1.3266883}
}

@article{germano1992,
  title={Turbulence: the filtering approach},
  author={Germano, Massimo},
  journal={Journal of Fluid Mechanics},
  volume={238},
  pages={325--336},
  year={1992},
  publisher={Cambridge University Press},
  doi={10.1017/S0022112092001733}
}

@article{kuzzay2017,
  title={New method for detecting singularities in experimental incompressible flows},
  author={Kuzzay, Denis and Saw, Ewe-Wei and Martins, Fabio JWA and Faranda, Davide and Foucaut, Jean-Marc and Daviaud, Francois and Dubrulle, Berengere},
  journal={Nonlinearity},
  volume={30},
  number={6},
  pages={2381},
  year={2017},
  publisher={IOP Publishing},
  doi={10.1088/1361-6544/aa6aaf}
}

@article{faranda2024,
  title={Statistical physics and dynamical systems perspectives on geophysical extreme events},
  author={Faranda, Davide and Messori, Gabriele and Alberti, Tommaso and Alvarez-Castro, Carmen and Caby, Th{\'e}ophile and Cavicchia, Leone and Coppola, Erika and Donner, Reik V and Dubrulle, Berengere and Galfi, Vera Melinda and others},
  journal={Physical Review E},
  volume={110},
  number={4},
  pages={041001},
  year={2024},
  publisher={APS},
  doi={10.1103/PhysRevE.110.041001}
}

@article{meneveau1994,
  title={Statistics of turbulence subgrid-scale stresses: Necessary conditions and experimental tests},
  author={Meneveau, Charles},
  journal={Physics of Fluids},
  volume={6},
  number={2},
  pages={815--833},
  year={1994},
  publisher={American Institute of Physics},
  doi={https://doi.org/10.1063/1.868320}
}

@article{storer2023,
  title={Global cascade of kinetic energy in the ocean and the atmospheric imprint},
  author={Storer, Benjamin A and Buzzicotti, Michele and Khatri, Hemant and Griffies, Stephen M and Aluie, Hussein},
  journal={Science Advances},
  volume={9},
  number={51},
  pages={eadi7420},
  year={2023},
  publisher={American Association for the Advancement of Science},
  doi={10.1126/sciadv.adi7420}
}

@incollection{leonard1975,
  title={Energy cascade in large-eddy simulations of turbulent fluid flows},
  author={Leonard, Anthony},
  booktitle={Advances in Geophysics},
  volume={18},
  pages={237--248},
  year={1975},
  publisher={Elsevier},
  doi={10.1016/S0065-2687(08)60464-1}
}

@article{best2011JULES,
AUTHOR = {Best, M. J. and Pryor, M. and Clark, D. B. and Rooney, G. G. and Essery, R. L. H. and M\'enard, C. B. and Edwards, J. M. and Hendry, M. A. and Porson, A. and Gedney, N. and Mercado, L. M. and Sitch, S. and Blyth, E. and Boucher, O. and Cox, P. M. and Grimmond, C. S. B. and Harding, R. J.},
TITLE = {The Joint UK Land Environment Simulator (JULES), model description – Part 1: Energy and water fluxes},
JOURNAL = {Geoscientific Model Development},
VOLUME = {4},
YEAR = {2011},
NUMBER = {3},
PAGES = {677--699},
URL = {https://gmd.copernicus.org/articles/4/677/2011/},
DOI = {10.5194/gmd-4-677-2011}
}

@article{donlon2012OSTIA,
  title={The operational sea surface temperature and sea ice analysis (OSTIA) system},
  author={Donlon, Craig J and Martin, Matthew and Stark, John and Roberts-Jones, Jonah and Fiedler, Emma and Wimmer, Werenfrid},
  journal={Remote sensing of Environment},
  volume={116},
  pages={140--158},
  year={2012},
  publisher={Elsevier}
}

@article{clark2011JULES,
AUTHOR = {Clark, D. B. and Mercado, L. M. and Sitch, S. and Jones, C. D. and Gedney, N. and Best, M. J. and Pryor, M. and Rooney, G. G. and Essery, R. L. H. and Blyth, E. and Boucher, O. and Harding, R. J. and Huntingford, C. and Cox, P. M.},
TITLE = {The Joint UK Land Environment Simulator (JULES), model description – Part 2: Carbon fluxes and vegetation dynamics},
JOURNAL = {Geoscientific Model Development},
VOLUME = {4},
YEAR = {2011},
NUMBER = {3},
PAGES = {701--722},
URL = {https://gmd.copernicus.org/articles/4/701/2011/},
DOI = {10.5194/gmd-4-701-2011}
}

@article{bush2025RAL3,
AUTHOR = {Bush, M. and Flack, D. L. A. and Lewis, H. W. and Bohnenstengel, S. I. and Short, C. J. and Franklin, C. and Lock, A. P. and Best, M. and Field, P. and McCabe, A. and Van Weverberg, K. and Berthou, S. and Boutle, I. and Brooke, J. K. and Cole, S. and Cooper, S. and Dow, G. and Edwards, J. and Finnenkoetter, A. and Furtado, K. and Halladay, K. and Hanley, K. and Hendry, M. A. and Hill, A. and Jayakumar, A. and Jones, R. W. and Lean, H. and Lee, J. C. K. and Malcolm, A. and Mittermaier, M. and Mohandas, S. and Moore, S. and Morcrette, C. and North, R. and Porson, A. and Rennie, S. and Roberts, N. and Roux, B. and Sanchez, C. and Su, C.-H. and Tucker, S. and Vosper, S. and Walters, D. and Warner, J. and Webster, S. and Weeks, M. and Wilkinson, J. and Whitall, M. and Williams, K. D. and Zhang, H.},
TITLE = {The third Met Office Unified Model--JULES Regional Atmosphere and Land Configuration, RAL3},
JOURNAL = {Geoscientific Model Development},
VOLUME = {18},
YEAR = {2025},
NUMBER = {12},
PAGES = {3819--3855},
URL = {https://gmd.copernicus.org/articles/18/3819/2025/},
DOI = {10.5194/gmd-18-3819-2025}
}

@article{walters2019GAL,
  title={The Met Office Unified Model global atmosphere 7.0/7.1 and JULES global land 7.0 configurations},
  author={Walters, David and Baran, Anthony J and Boutle, Ian and Brooks, Malcolm and Earnshaw, Paul and Edwards, John and Furtado, Kalli and Hill, Peter and Lock, Adrian and Manners, James and others},
  journal={Geoscientific Model Development},
  volume={12},
  number={5},
  pages={1909--1963},
  year={2019},
  publisher={Copernicus GmbH},
  doi={10.5194/gmd-12-1909-2019}
}

@article{field2023CASIM,
  title={Implementation of a double moment cloud microphysics scheme in the UK met office regional numerical weather prediction model},
  author={Field, Paul R and Hill, Adrian and Shipway, Ben and Furtado, Kalli and Wilkinson, Jonathan and Miltenberger, Annette and Gordon, Hamish and Grosvenor, Daniel P and Stevens, Robin and Van Weverberg, Kwinten},
  journal={Quarterly Journal of the Royal Meteorological Society},
  volume={149},
  number={752},
  pages={703--739},
  year={2023},
  publisher={Wiley Online Library},
  doi={10.1002/qj.4414}
}

@article{wilsonballard1999,
  title={A microphysically based precipitation scheme for the UK Meteorological Office Unified Model},
  author={Wilson, Damian R and Ballard, Susan P},
  journal={Quarterly Journal of the Royal Meteorological Society},
  volume={125},
  number={557},
  pages={1607--1636},
  year={1999},
  publisher={Wiley Online Library},
  doi={10.1002/qj.49712555707}
}

@article{KVW2021bimodal,
  title={A bimodal diagnostic cloud fraction parameterization. Part I: Motivating analysis and scheme description},
  author={Van Weverberg, Kwinten and Morcrette, Cyril J and Boutle, Ian and Furtado, Kalli and Field, Paul R},
  journal={Monthly Weather Review},
  volume={149},
  number={3},
  pages={841--857},
  year={2021},
  doi={MWR-D-20-0224.1}
}

@article{wilson2008pc2,
  title={PC2: A prognostic cloud fraction and condensation scheme. I: Scheme description},
  author={Wilson, Damian R and Bushell, Andrew C and Kerr-Munslow, Amanda M and Price, Jeremy D and Morcrette, Cyril J},
  journal={Quarterly Journal of the Royal Meteorological Society},
  volume={134},
  number={637},
  pages={2093--2107},
  year={2008},
  publisher={Wiley Online Library},
  doi={doi/10.1002/qj.333}
}

@article{brown2012UM,
  title={Unified modeling and prediction of weather and climate: A 25-year journey},
  author={Brown, Andrew and Milton, Sean and Cullen, Mike and Golding, Brian and Mitchell, John and Shelly, Ann},
  journal={Bulletin of the American Meteorological Society},
  volume={93},
  number={12},
  pages={1865--1877},
  year={2012},
  publisher={American Meteorological Society},
  doi={10.1175/BAMS-D-12-00018.1}
}

@article{y07p1,
  title={Convectively coupled equatorial waves. Part I: Horizontal and vertical structures},
  author={Yang, Gui-Ying and Hoskins, Brian and Slingo, Julia},
  journal={Journal of the Atmospheric Sciences},
  volume={64},
  number={10},
  pages={3406--3423},
  year={2007},
  doi={10.1175/JAS4017.1}
}

@article{SK03,
  title={The observed structure of convectively coupled Kelvin waves: Comparison with simple models of coupled wave instability},
  author={Straub, Katherine H and Kiladis, George N},
  journal={Journal of the Atmospheric Sciences},
  volume={60},
  number={14},
  pages={1655--1668},
  year={2003},
  publisher={American Meteorological Society},
  doi={10.1175/1520-0469(2003)060<1655:TOSOCC>2.0.CO;2}
}

@article{NARVAL,
  title={A high-altitude long-range aircraft configured as a cloud observatory: The NARVAL expeditions},
  author={Stevens, Bjorn and Ament, Felix and Bony, Sandrine and Crewell, Susanne and Ewald, Florian and Gross, Silke and Hansen, Akio and Hirsch, Lutz and Jacob, Marek and K{\"o}lling, Tobias and others},
  journal={Bulletin of the American Meteorological Society},
  volume={100},
  number={6},
  pages={1061--1077},
  year={2019},
  publisher={American Meteorological Society},
  doi={10.1175/BAMS-D-18-0198.1}
}

@article{roundy2008,
  title={Analysis of convectively coupled Kelvin waves in the Indian Ocean MJO},
  author={Roundy, Paul E},
  journal={Journal of the Atmospheric Sciences},
  volume={65},
  number={4},
  pages={1342--1359},
  year={2008},
  publisher={American Meteorological Society},
  doi={10.1175/2007JAS2345.1}
}

@article{tomassini2023,
author = {Tomassini, Lorenzo and Willett, Martin and Sellar, Alistair and Lock, Adrian and Walters, David and Whitall, Michael and Sanchez, Claudio and Heming, Julian and Earnshaw, Paul and Rodriguez, José M. and Ackerley, Duncan and Xavier, Prince and Franklin, Charmaine and Senior, Catherine A.},
title = {Confronting the Convective Gray Zone in the Global Configuration of the Met Office Unified Model},
journal = {Journal of Advances in Modeling Earth Systems},
volume = {15},
number = {5},
pages = {e2022MS003418},
doi = {https://doi.org/10.1029/2022MS003418},
year = {2023}
}

@article {NastromGage1985,
      author = "G. D.  Nastrom and K. S.  Gage",
      title = "A Climatology of Atmospheric Wavenumber Spectra of Wind and Temperature Observed by Commercial Aircraft",
      journal = "Journal of Atmospheric Sciences",
      year = "1985",
      publisher = "American Meteorological Society",
      address = "Boston MA, USA",
      volume = "42",
      number = "9",
      doi = "10.1175/1520-0469(1985)042<0950:ACOAWS>2.0.CO;2",
      pages=      "950 - 960",
      url = "https://journals.ametsoc.org/view/journals/atsc/42/9/1520-0469_1985_042_0950_acoaws_2_0_co_2.xml"
}

@article{Stephan2022,
 author = {Stephan, Claudia Christine and Duras, Julia and Harris, Lucas and Klocke, Daniel and Putman, William M. and Taylor, Mark and Wedi, Nils P. and Žagar, Nedjeljka and Ziemen, Florian},
 journal = {Tellus A: Dynamic Meteorology and Oceanography},
 month = {Apr},
 title = {Atmospheric Energy Spectra in Global Kilometre-Scale Models},
 year = {2022},
 doi = {10.16993/tellusa.26}
}

@article{clayton2013UMA,
author = {Clayton, A. M. and Lorenc, A. C. and Barker, D. M.},
title = {Operational implementation of a hybrid ensemble/4D-Var global data assimilation system at the Met Office},
journal = {Quarterly Journal of the Royal Meteorological Society},
volume = {139},
number = {675},
pages = {1445-1461},
doi = {https://doi.org/10.1002/qj.2054},
year = {2013}
}

@article{DaSilvaGPM,
author = {Da Silva, Nicolas A. and Webber, Benjamin G. M. and Matthews, Adrian J. and Feist, Matthew M. and Stein, Thorwald H. M. and Holloway, Christopher E. and Abdullah, Muhammad F. A. B.},
title = {Validation of GPM IMERG Extreme Precipitation in the Maritime Continent by Station and Radar Data},
journal = {Earth and Space Science},
volume = {8},
number = {7},
pages = {e2021EA001738},
doi = {10.1029/2021EA001738},
year = {2021}
}

@software{LoSSETT,
  author       = {Shipley, D. and McKinnon-Gray, E.},
  title        = {LoSSETT (Local Scale-to-Scale Energy Transfer Tool)},
  url          = {https://github.com/ElliotMG/LoSSETT},
  version      = {v0.2},
  date         = {2025-06-30},
  year         = {2025}
}

@article{Dee2011ERAI,
author = {Dee, D. P. and Uppala, S. M. and Simmons, A. J. and Berrisford, P. and Poli, P. and Kobayashi, S. and Andrae, U. and Balmaseda, M. A. and Balsamo, G. and Bauer, P. and Bechtold, P. and Beljaars, A. C. M. and van de Berg, L. and Bidlot, J. and Bormann, N. and Delsol, C. and Dragani, R. and Fuentes, M. and Geer, A. J. and Haimberger, L. and Healy, S. B. and Hersbach, H. and Hólm, E. V. and Isaksen, L. and Kållberg, P. and Köhler, M. and Matricardi, M. and McNally, A. P. and Monge-Sanz, B. M. and Morcrette, J.-J. and Park, B.-K. and Peubey, C. and de Rosnay, P. and Tavolato, C. and Thépaut, J.-N. and Vitart, F.},
title = {The ERA-Interim reanalysis: configuration and performance of the data assimilation system},
journal = {Quarterly Journal of the Royal Meteorological Society},
volume = {137},
number = {656},
pages = {553-597},
doi = {https://doi.org/10.1002/qj.828},
url = {https://rmets.onlinelibrary.wiley.com/doi/abs/10.1002/qj.828},
year = {2011}
}

@software{waves,
    author  =   {McKinnon-Gray, E.},
    title   =   {realtime\_waves},
    url     =   {https://github.com/ElliotMG/realtime_waves},
    date    =   {2025-11-06},
    year    =   {2025}
}

@article{JRB21,
author = {Judt, Falko and Rios-Berrios, Rosimar},
title = {Resolved Convection Improves the Representation of Equatorial Waves and Tropical Rainfall Variability in a Global Nonhydrostatic Model},
journal = {Geophysical Research Letters},
volume = {48},
number = {14},
pages = {e2021GL093265},
doi = {10.1029/2021GL093265},
year = {2021}
}

@article{Vosper2016,
author = {Vosper, S. B. and Brown, A. R. and Webster, S.},
title = {Orographic drag on islands in the NWP mountain grey zone},
journal = {Quarterly Journal of the Royal Meteorological Society},
volume = {142},
number = {701},
pages = {3128-3137},
doi = {10.1002/qj.2894},
year = {2016}
}

@article{Johnson2020,
  title = {Energy Transfer from Large to Small Scales in Turbulence by Multiscale Nonlinear Strain and Vorticity Interactions},
  author = {Johnson, Perry L.},
  journal = {Phys. Rev. Lett.},
  volume = {124},
  issue = {10},
  pages = {104501},
  numpages = {6},
  year = {2020},
  month = {Mar},
  publisher = {American Physical Society},
  doi = {10.1103/PhysRevLett.124.104501},
  url = {https://link.aps.org/doi/10.1103/PhysRevLett.124.104501}
}

\end{document}